\documentclass[11pt,preprint]{aastex}
\usepackage{graphicx}
\usepackage{amsmath} 
\usepackage{subfigure}
\bibpunct{(}{)}{;}{a}{,}{,}
\begin{document}


\def\etal{et al.\ \rm}
\def\ba{\begin{eqnarray}}
\def\ea{\end{eqnarray}}

\title{Non-Power Law Behavior in Fragmentation Cascades}

\author{Mikhail A. Belyaev\altaffilmark{1} \& 
Roman R. Rafikov\altaffilmark{1,2}}
\altaffiltext{1}{Department of Astrophysical Sciences, 
Princeton University, Ivy Lane, Princeton, NJ 08540; 
mbelyaev@astro.princeton.edu, rrr@astro.princeton.edu}
\altaffiltext{2}{Sloan Fellow}


\begin{abstract}
Collisions resulting in fragmentation are important in shaping 
the mass spectrum of minor bodies in the asteroid belt, the Kuiper belt, 
and debris disks. Models of fragmentation cascades typically find
that in steady-state, the solution for the particle mass distribution
is a power law in the mass. However, previous studies have typically 
assumed that the mass of the largest fragment produced in a collision
with just enough energy to shatter the target and disperse half its
mass to infinity
is directly proportional to the target mass. We show 
that if this assumption is not satisfied, then the power law solution 
for the steady-state particle mass distribution is modified by 
a multiplicative factor, which is a slowly varying function of the 
mass. We derive analytic solutions for this correction factor and confirm our
results numerically. We find that this correction factor proves
important when extrapolating over many orders of magnitude in mass,
such as when inferring the number of large objects in a
system based on infrared observations. In the course of our work, we
have also discovered an unrelated type of non-power law behavior: waves
can persist in the mass distribution of objects even in the absence of upper or
lower cutoffs to the mass distribution or breaks in the strength law.
\end{abstract}



\section{Introduction.}  
\label{sect:intro}


Collisional evolution of many-body astrophysical systems in which 
the relative velocity between colliding objects is large compared 
to the escape speed and coagulation is unimportant 
is dominated by fragmentation. Examples of such systems include the
asteroid belt \citep{DD, Bottke}, the Kuiper Belt
\citep{DavisFarinella, PanSari}, and
debris disks around young stars \citep{KW,KB}. As the large bodies 
are slowly ground down, a collisional cascade is 
launched, in which mass flows unidirectionally to smaller 
objects until at some scale it is flushed out of the system by 
some removal process (e.g. by Poynting-Robertson drag, radiation 
pressure, or gas drag). 

In real astrophysical systems there is normally a large dynamic 
range between the scale $m_{inj}$ at which mass is injected into 
the cascade and the scale $m_{rm}$ at which mass is removed 
from the system.  The mass $m_{inj}$ can be defined as
the mass for which the collisional destruction timescale is 
comparable to the age of the system. In highly evolved systems, such
as the Kuiper Belt, this scale corresponds to the characteristic 
mass of the largest bodies \citep{PanSari, Fraser}.
The collision time for the smallest bodies is usually much shorter 
than for the largest ones, so a steady-state can be set up
for $m_{rm}\ll m\ll m_{inj}$. 
In such a steady-state, the mass distribution of particles evolves 
on the collision timescale of bodies at the injection mass scale 
and can be considered static on shorter timescales. 

\citet{Dohnanyi} was the first to construct a model of a 
fragmentation cascade that aimed at explaining the mass 
distribution of objects in the asteroid belt. He assumed that the
internal strength of colliding objects is independent of mass 
with the implication that $m_B(m_t)$, which we define as the mass of
the smallest projectile capable of
dispersing one half the mass of a target of mass $m_t$ to
infinity, has its mass
linearly proportional to $m_t$. Dohnanyi also assumed that $m_0(m_t,m_p)$,
the mass of the largest fragment produced in a collision between a
target of mass $m_t$ and a projectile of mass $m_p$, 
scales linearly with $m_t$ and is independent of $m_p$. Under these
assumptions, he showed
that a fragmentation cascade allows a steady-state power law solution 
\ba
n(m)\propto m^{-\alpha}, ~~~\alpha = -11/6
\label{eq:PL}
\ea
where $n(m)dm$ is the number of objects in the size distribution with
mass between $m$ and $m + dm$.

Later, \citet{Tanaka} generalized Dohnanyi's result by assuming
a self-similar model of fragmentation, which again entails
$m_B(m_t)\propto m_t$, but now $m_0(m_t,m_p) \propto m_t q(m_p/m_t)$,
where $q$ is an arbitrary function. They confirmed the power
law form of the mass spectrum and showed that the value of $\alpha$
is determined by the mass dependence of the collision rate and that
$\alpha$ reduces to $11/6$ if the collision rate is proportional to
$m_t^{2/3}$ (geometrical cross-section with mass-independent relative
velocities).

In reality, however, fragmentation does not have to be self-similar 
because the minimum energy necessary for disruption is not always 
linearly proportional 
to the target mass. For instance, if an object's internal strength is
dominated by gravity, then $Q_D^\star$, the energy per unit mass required to
shatter an object and disperse half of its mass to infinity,
 scales as $m_t Q_D^\star \propto
m_t^{5/3}$, which is the case for objects larger than $\sim 1$ km
\citep{BenzAsphaug, Holsapple, BAR}. \citet{OBG} considered a model in which 
$Q_D^\star$ scales as a power law in
$m_t$, but took $m_0$ linearly proportional to $m_t$.
Under these assumptions, \citet{OBG} found that a steady-state 
power law solution for $n(m)$ still exists, but that 
$\alpha$ differs from $11/6$ unless $Q_D^\star$ is constant, even when
the collision rate scales as $m_t^{2/3}$.

The steady-state power law solutions of \citet{Dohnanyi},
\citet{Tanaka}, and \citet{OBG}
have since been confirmed numerous times by simulations, which have
also shed light on non-power law effects present in astrophysical fragmentation
cascades. These effects typically manifest themselves as waves
superimposed on top of the steady-state power law solution and are
caused either by non-collisional mass sinks, e.g. due to the removal of
small particles ($\lesssim 1$ $\mu$m for $L=L_\sun$)
by radiation pressure \citep{Thebault, CampoBagatin, DD}; a change in
the power
law index of $Q_D^\star(m_t)$ induced by a transition from the
strength-dominated to the gravity-dominated regime
\citep{OBG, OBG1}; or a transition from a primordial to a
collisionally-evolved size distribution \citep{Fraser, PanSari, KB2004}.

In this work, we describe a new source of non-power law behavior
in fragmentation cascades. We consider a model similar to the one 
used by \citet{OBG}, but with a more general form for $m_0$. More
specifically, previous researchers \citep{PetitFarinella, OBG1,
  Williams, KobayashiTanaka, Brunini} have typically assumed that $m_0
= m_t p(E_{coll}/m_tQ_D^\star)$, where
$E_{coll}$ is the kinetic energy of the colliding particles in the
center of mass frame, and $p$ is a function that varies from author to
author. We instead consider the more general dependence
$m_0 = m_t^\mu p(E_{coll}/m_tQ_D^\star)$, which is motivated in \S
\ref{setup}. We find that unless $\mu=1$,
there is no steady-state power law solution. Instead, $n(m)$ is described by a
power law with the same power law index as found by \citet{OBG}, but
multiplied by a slowly varying function of
mass (i.e. $n(m) \propto m^{-\alpha} \varphi(m), ~~~|d \ln \varphi/d
\ln m| \ll 1$). The non-power law effects caused by the slowly varying
function show up as a
smooth deviation from power-law behavior, which is quite different from
the wave-like features described earlier. This deviation
is significant when extrapolating over many orders of magnitude in
mass, as demonstrated in \S\ref{subsect:applications}.

In the course of our investigation, we have also discovered that it is possible
for waves to appear and persist in a collisional cascade, even if the
particle size
distribution does not contain an upper or lower mass cutoff, and the
strength is described by a pure power law with no breaks. This is a
completely independent type of non-power law behavior from the one
caused by $m_0$ not proportional to $m_t$. In astrophysical systems,
such waves may be triggered by stochastic collision events between
large planetesimals \citep{KB2005, WyattDent,
  Wyatt}. However, the main focus of our paper is on the
non-power law behavior that results when $m_0$ is not proportional to
$m_t$, and we defer a detailed discussion of these waves for the future.

The paper is organized as follows. In \S\ref{setup}, we introduce the
general equations describing 
fragmentation and discuss specific assumptions relevant to our
model. In \S \ref{pls} we demonstrate that 
pure power law solutions for a fragmentation cascade are indeed 
possible if $m_0 \propto m_t p(E_{coll}/m_tQ_D^\star)$, consistent
with \citet{KobayashiTanaka}.
We show in \S \ref{nonpl} that if the assumption $m_0 \propto m_t$ 
is not satisfied, then solutions are given by the product of a
power law with the same index as in the $m_0 \propto m_t$ case, 
and a slowly varying function of mass. We then find
analytic solutions for this slowly varying function for monodisperse
(all fragments having the same mass) and power law fragment mass
distributions. We find that in the monodisperse case, the solutions
for the steady-state distribution are not unique and can support
waves. We confirm our analytical results numerically in \S
\ref{numerical} and discuss their validity and possible applications
in \S \ref{discussion}. 
    

\section{Basic setup}
\label{setup}


The number density distribution of particles in mass space, $n(m)$, 
obeys the continuity equation \citep{Tanaka}:
\ba
\frac{\partial n(m)m}{\partial t}+\frac{\partial F(m)}{\partial m}=0,
\label{eq:continuity}
\ea
where the mass flux, $F(m)$ is defined to be the amount of mass
which flows past a point $m$ in mass space per unit time. We will
consider the evolution of the mass spectrum for $m\ll m_{inj}$
in which case we can adopt the steady-state assumption.
Under this assumption, $F(m)$ is constant, so there is no accumulation of
particles at any scale.

In order to write down the explicit form of $F(m)$, we 
need to make some definitions. Disruption of a target in a collision
produces a spectrum
of fragments characterized by the function $g(m_f,m_t,m_p)$, where 
$g(m_f,m_t,m_p)dm_f$ is the number of fragments in the mass interval 
$(m_f,m_f+dm_f)$ coming {\it only from the target} in a collision 
between bodies with mass $m_p$ and $m_t$. Mass conservation then requires that
\ba
\int\limits_0^{\infty}g(m_f,m_t,m_p)m_f dm_f=m_t.
\label{eq:norm}
\ea

\citet{KobayashiTanaka} have shown that erosive collisions provide
the dominant contribution to the mass flux, and in order to take
them into account, it is useful to split $g$ into two components
\ba
g(m_f,m_t,m_p) = g_{rem}(m_f,m_t,m_p) + g_{ej}(m_f,m_t,m_p). 
\ea 
Here, $g_{ej}$ is the contribution to the mass flux from the
continuous distribution of fragments ejected from the target and is
normalized such that
\ba
\int\limits_0^{\infty}g_{ej}(m_f,m_t,m_p)m_f dm_f=m_{ej}(m_t,m_p),
\label{eq:gej}
\ea
where $m_{ej}$ is the total amount of mass ejected from the target and
dispersed to infinity. In catastrophic
collisions with $m_p \gg m_B$, $m_{ej} = m_t$. However, in collisions with $m_p
\lesssim m_B$, the core of the target is left almost intact \citep{Fujiwara},
leaving behind a remnant of mass $m_{rem} = m_t - m_{ej}$. This yields
\ba
g_{rem}(m_f,m_t,m_p) = \delta(m_f-m_{rem}(m_t,m_p)).
\label{eq:grem}
\ea
When $m_p = m_B$, then by definition of $m_B$, $m_{rem} = m_{ej} =
m_t/2$. A collision is
commonly referred to as catastrophic when $m_{ej} > m_{rem}$ and as erosive
when $m_{ej} < m_{rem}$. We also clarify that $m_0$ is the largest
fragment in the continuous distribution of ejecta, $g_{ej}$, to which $m_{rem}$
does {\it not} belong.

We next define $f(m,m_t,m_p)$ as the mass fraction of debris 
with $m_f<m$, which comes from the target only 
in a collision between a target of mass $m_t$ and a projectile 
of mass $m_p$: 
\ba
f(m,m_t,m_p)=\frac{1}{m_t}
\int\limits_0^{m}g(m_f,m_t,m_p)m_f dm_f.
\label{eq:gdef}
\ea
This is consistent with
\citet{Tanaka} and \citet{KobayashiTanaka} and allows for projectiles
that are larger
than targets. We next split $f$ into two parts, just as we did with
$g$: 
\ba
f(m,m_t,m_p) = f_{ej}(m,m_t,m_p) + f_{rem}(m,m_t,m_p).
\label{fsplit}
\ea
Using the definitions (\ref{eq:gej}), (\ref{eq:grem}),
and (\ref{eq:gdef}), we have
\ba
f_{ej}(m,m_t,m_p) =
\frac{1}{m_t}\int\limits_0^{m}g_{ej}(m_f,m_t,m_p)m_f dm_f,
\label{eq:fej}
\ea
and
\ba
f_{rem}(m,m_t,m_p) = \left \{ 
\begin{array}{lr}
  m_{rem}(m_t,m_p)/m_t &,~~~m > m_{rem}(m_t,m_p) \\
  0 &, ~~~m < m_{rem}(m_t,m_p) 
\end{array}.
\right.
\ea

We can now write down the mass flux in the form
\begin{equation}
\label{massflux}
F(m) = -\int\limits_m^\infty dm_t m_t n(m_t) 
\int\limits_{0}^{\infty} dm_p
f(m,m_t,m_p)n(m_p){\cal R}(m_t,m_p),
\end{equation}
where ${\cal R}(m_t, m_p)$ 
is the collision rate between bodies with mass $m_t$ and 
$m_p$. The utility of splitting $f$ into two components (Eq. \ref{fsplit})
will become apparent shortly in \S\ref{fragmentationmodel}.


\subsection{Some simplifications}
\label{simplifications}

When gravitational focusing is unimportant, the collision rate is
given by
\ba
{\cal R}(m_t,m_p) = \pi \left(\frac{3}{4 \pi \rho}\right)^{2/3}
\left(m_t^{1/3}+m_p^{1/3}\right)^2 \langle v \rangle^2,
\label{eq:coll_rate_gen0}
\ea
where $\langle v \rangle$ is an averaged collision velocity, which can
be a function of $m_t$ and $m_p$. To limit the number of parameters in
our study, we will assume 
\ba
\langle v \rangle = const,
\label{eq:vconstant}
\ea
so the collision rate becomes 
\ba
{\cal R}(m_t,m_p) \propto \left(m_t^{1/3}+m_p^{1/3}\right)^2,
\label{eq:coll_rate_gen}
\ea
but our results are easily extended to the forms of ${\cal R}$
considered by \citet{Tanaka}, who varied
the power law dependence of ${\cal R}$ on $m_t$ and $m_p$.

It is natural to expect that disruption of targets with mass $m_t$
is dominated by collisions with projectiles having masses near or
below the 
breaking threshold $m_B(m_t)$. This is because the cross-section for
catastrophic collisions (defined by $m_p > m_B(m_t)$ (\S\ref{setup}))
is dominated by the smallest particles as long
as $\alpha > 5/3$ \citep{Dohnanyi}, and the mass flux from erosive
collisions drops off for $m_p \ll m_B(m_t)$
\citep{KobayashiTanaka}. We now assume
\ba 
m_B(m_t) \ll m_t,
\label{eq:mBllmt}
\ea 
which is typically valid for astrophysical fragmentation
cascades. Then, ${\cal R}(m_t)\propto
m_t^{2/3}$ for collisions that are responsible for the
majority of the mass flux.  This allows us to rewrite (\ref{massflux})
in the following form:
\begin{equation}
\label{massflux1}
F(m) \propto \int\limits_m^\infty dm_t m_t n(m_t){\cal R}(m_t) 
\int\limits_{0}^{\infty} dm_p f(m,m_t,m_p) n(m_p).
\end{equation}
The normalization of $F(m)$ is unimportant since the only thing that
matters for a steady-state solution is that $F(m)$ is constant.


\subsection{Fragmentation Model}
\label{fragmentationmodel}

Experimental data on collisional breakup \citep{Gault,Fujiwara}
and numerical simulations of high-velocity collisions 
\citep{BAR,BenzAsphaug} suggest that the
mass spectrum of fragments ejected from the target in a single collision
can be reasonably well fit within a broad range of masses by a  
power-law with a cutoff at $m_0(m_t,m_p)$\footnote{\citet{Fujiwara},
\citet{Takagi}, \citet{Davis} find that a two or three slope power law
better fits the data.}:
\ba
g_{ej}(m_f,m_t,m_p)\propto \left \{ 
\begin{array}{lr}
  m_f^{-\eta(m_t,m_p)} &,~~~m_f < m_0(m_t,m_p) \\
  0 &, ~~~m_f > m_0(m_t,m_p)
\end{array}
\right.
\label{eq:fragPL}
\ea

Equation (\ref{eq:fragPL}) is difficult to analyze for arbitrary
dependencies of $\eta(m_t,m_p)$ and $m_0(m_t,m_p)$. Thus, we make the
simplification, motivated in \S\ref{subsect:m0_mt} that
$g_{ej}(m_f,m_t,m_p)$ has the form\footnote{In
  some sense, this is more general than the form (\ref{eq:fragPL}),
  because $g_{ej}$ does not have to be a power law.}
\ba
g_{ej}(m_f,m_t,m_p)=\frac{m_{ej}(m_t,m_p)}{m_{0,B}^2(m_t)}\psi\left(\frac{m_f}
{m_{0,B}(m_t)},\frac{E_{coll}(m_t,m_p)}{m_tQ_D^\star(m_t)}\right),
~~~m_{0,B}(m_t) \equiv m_0(m_t,m_B(m_t)).
\label{eq:psi}
\ea
In the limit $m_p \ll m_t$,
\ba
E_{coll}(m_t,m_p) \approx m_p\langle v \rangle^2/2,
\label{eq:eclim}
\ea
which together with the assumptions (\ref{eq:mBllmt}) and the
definition\footnote{Some authors
  (e.g. \citet{OBG1}) assume that the projectile absorbs half of the
  energy, so $E_{coll}(m_t,m_B(m_t)) = 2Q_D^\star m_t$, but it does not
  matter which definition is adopted for our
  purposes.} 
\ba
E_{coll}(m_t,m_B(m_t)) = Q_D^\star(m_t)m_t
\label{eq:Ecrit}
\ea
implies
\ba
\frac{E_{coll}(m_t,m_p)}{m_tQ_D^\star(m_t)} =
\frac{m_p}{m_B(m_t)}, ~~~ m_B(m_t) \ll m_t.
\label{eq:Ecoll} 
\ea

It now helps to define the variables 
\ba
\label{eq:x}
x &\equiv& m/m_{0,B}(m_t) \\
\label{eq:y}
y &\equiv& m_p/m_B(m_t) \\
\label{eq:z}
z &\equiv& m/m_t
\ea 
Then, with the form of $g_{ej}$ given by Eq. (\ref{eq:psi}), $f_{ej}$ becomes
\ba
f_{ej}(m,m_t,m_p) = \frac{m_{ej}(m_t,m_p)}{m_t}\xi(x,y).
\label{eq:fredef}
\ea
The normalization of $\xi$ is such that $\xi(\infty, y) = 1$, for
$y>0$, which follows from Eq. (\ref{eq:gej}) and Eq. (\ref{eq:fej}).

The prefactor in Eq. (\ref{eq:fredef}) can be written as
\ba
\frac{m_{ej}(m_t,m_p)}{m_t} = 1 - \frac{m_{rem}(m_t,m_p)}{m_t}.
\ea
If we now make the assumption that 
\ba
\frac{m_{rem}(m_t,m_p)}{m_t} =
\chi(y),
\label{eq:rem} 
\ea
we can absorb the prefactor in Eq. (\ref{eq:fredef}) and write
\ba
f_{ej}(m,m_t,m_p) = f_{ej}(x,y).
\ea
One can consider more general prescriptions for $m_{rem}/m_t$ than
Eq. (\ref{eq:rem}), but the
latter suffices to illustrate the non-power law behavior. It also
follows from Eq. (\ref{eq:rem}) that $f_{rem}$ has the form
\ba
f_{rem}(z,y)
= \left \{ 
\begin{array}{lr}
  \chi(y)
  &,~~~z > \chi(y) \\
  0 &, ~~~z < \chi(y)
\end{array},
\right.
\ea
so that finally we arrive at
\ba
f(m,m_t,m_p) =
f_{rem}(z,y) + f_{ej}(x,y).
\label{eq:fcombo}
\ea


\subsection{Nonlinear scaling of $m_{0,B}(m_t)$.}
\label{subsect:m0_mt}

We now address the natural question of whether one should 
expect a nonlinear scaling of $m_{0,B}(m_t)$ in practice? 
Experimental \citep{Fujiwara} and numerical  
\citep{BenzAsphaug} studies of collisional fragmentation 
suggest that the mass of the largest fragment 
formed in a high-speed catastrophic collision decreases with
increasing collision energy. In particular, we focus on the
experiments of \citet{Fujiwara}, who fired polycarbonate projectiles
of a constant mass and kinetic energy ($m_p = .37$ g $v_p = 2.6$
km $\text{s}^{-1}$) into basalt targets with masses in the range $22 \
\text{g} < m_t < 2900 \ \text{g}$. They fit their data in the
catastrophic regime with the following relation:
\ba
\max[m_0(m_t|m_p,v_p),m_{rem}(m_t|m_p,v_p)] \propto
m_t
\left(\frac{E_{coll}(m_t|m_p,v_p)}{m_t}\right)^{-\gamma},~~~\gamma\approx
1.24.
\label{eq:Fujiw}
\ea
The vertical bar denotes the fact that the experiments of \citet{Fujiwara}
were performed at constant projectile mass and
velocity. Since $E_{coll}(m_t|m_p,v_p) \approx const$ for $m_p \ll m_t$,
what \citet{Fujiwara}
have shown is that $m_0(m_t|m_p,v_p) \propto m_t^{1+\gamma}$. However,
we now make the following extension to their results. We assume
Eq. (\ref{eq:Fujiw}) to be valid as a function of $m_p$ as well, so
we write 
\ba
\max[m_0(m_t,m_p|v_p),m_{rem}(m_t,m_p|v_p)] \propto
m_t \left(\frac{E_{coll}(m_t,m_p|v_p)}{m_t}\right)^{-\gamma}
\label{eq:Fujiw2}
\ea
In the highly catastrophic fragmentation regime ($E_{coll} \gg
m_tQ_D^*$), we expect all of the fragments to be a part of the
continuous fragment distribution, so that there is no remnant mass
remaining ($m_{rem} = 0$). This means we can simplify
Eq. (\ref{eq:Fujiw2}) to the form
\ba
m_0(m_t,m_p|v_p) \propto m_t
\left(\frac{E_{coll}(m_t,m_p|v_p)}{m_t}\right)^{-\gamma}.
\label{eq:Fujiw3}
\ea
Using $m_p = m_B(m_t)$ in Eq. (\ref{eq:eclim}) and making the usual
assumption (\ref{eq:vconstant}) then yields
\ba
\frac{m_{0,B}(m_t)}{m_t} \propto
\left(\frac{m_B(m_t)}{m_t}\right)^{-\gamma}.
\label{eq:m0BFuj}
\ea
Thus, unless $m_B \propto m_t$, $m_{0,B}$ is
{\it not} proportional to $m_t$. Moreover, we see that if $m_B$ varies
as a power law, then $m_{0,B}$ varies as a power law as
well. One assumption that we have made in deriving Eq. (\ref{eq:m0BFuj})
is that Eq. (\ref{eq:Fujiw3}) is valid for $m_0 \lesssim m_t$, even
though \citet{Fujiwara} obtained the power law relationship
(Eq. (\ref{eq:Fujiw})) by fitting primarily to data in the regime
$m_0 \ll m_t$. Nevertheless, our analysis has shown that a power law
dependence for $m_{0,B}$ is plausible, and we discuss the matter further in \S
\ref{subsect:compare}.

We now deduce what we would expect
for the power law exponent of $m_{0,B}(m_t)$ in the range of target
masses considered by \citet{Fujiwara}. Experiments and simulations
\citep{BenzAsphaug, Holsapple, Housen, BAR}
show that $Q_D^*$ is well-described over a large range of masses by
the expression
\ba
Q_D^*(m_t) = Q_0 m_t^{s/3},
\label{eq:QDstar}
\ea
where the exponent $s$ is different for the strength-dominated and
gravity-dominated regimes.
If $\langle v \rangle$ is constant as we have been assuming, then 
it follows from Eq. (\ref{eq:Ecoll}) that
\ba
m_B(m_t) = B m_t^\beta, ~~~ \beta = 1 + s/3,
\label{eq:mB}
\ea
and consequently
\ba
m_{0,B}(m_t) = Cm_t^{\mu}, \ \mu = 1-\gamma s/3.
\label{eq:m0B}
\ea
From simulations of impacts into basalt with $v_p = 3$ km s$^{-1}$,
\citet{BenzAsphaug} found that $s = -.38$ in the strength-dominated
regime. Using $\gamma=1.24$, the value measured by \citet{Fujiwara} yields $\mu =
1.16$. This is a small deviation from $m_{0,B} \propto m_t$
(i.e. $\mu = 1$), but
enough to cause noticeable effects for real systems as we demonstrate
in \S \ref{subsect:applications}.

We next motivate our form for $g_{ej}$ in \S\ref{fragmentationmodel} by
assuming the more general form
\ba
g_{ej}(m_f,m_t,m_p)=\frac{m_{ej}(m_t,m_p)}{m_{0}^2(m_t,m_p)}\psi_1\left(\frac{m_f}
{m_0(m_t,m_p)},y\right)
\label{eq:gejgen}
\ea
and showing that it reduces to Eq. (\ref{eq:psi}) if $m_0(m_t,m_p)$ is
given by Eq. (\ref{eq:Fujiw3}) (with the assumption $v_p = const$). Using Eq. (\ref{eq:eclim}) and
Eq. ({\ref{eq:m0BFuj}}) in Eq. (\ref{eq:Fujiw3}), we can write 
\ba
m_0(m_t,m_p) \propto m_{0,B}(m_t) \left(\frac{m_p}{m_B(m_t)}\right)^{-\gamma}.
\ea
Substituting this expression into Eq. (\ref{eq:gejgen}) and using the
definition of $y$ from Eq. (\ref{eq:y}), yields
\ba
g_{ej}(m_f,m_t,m_p) =
\frac{m_{ej}(m_t,m_p)}{\left(m_{0,B}(m_t)
  y^{-\gamma}\right)^2}\psi_1\left(\frac{m_f}{m_{0,B}(m_t)}y^\gamma,y\right).
\ea
Making the definition $\psi(x,y) = y^{2\gamma}\psi_1(xy^\gamma,y)$, we
arrive at Eq. (\ref{eq:psi}).


\section{Power Law Solutions}
\label{pls}


We now look for a steady-state power law solution for the mass
distribution. Plugging Eq. (\ref{eq:PL}) into Eq. (\ref{massflux1}),
using ${\cal R}(m_t) \propto m_t^{2/3}$, and using the form of $f$
given in Eq. (\ref{eq:fcombo}) yields
\ba
F(m)\propto \int\limits_m^\infty dm_t m_t^{5/3-\alpha} \int\limits_0^\infty
dm_p m_p^{-\alpha} \left(f_{rem}(z,y) + f_{ej}(x,y) \right).
\label{eq:flux5}
\ea 

We now demonstrate how the power law solutions previously derived in the 
literature follow from this equation 
and elucidate under what conditions they fail.


\subsection{Dohnanyi (1969) and Tanaka \etal (1996) case}
\label{subsect:Dohnanyi}

\citet{Dohnanyi} and \citet{Tanaka} assumed a scale-free
model of fragmentation with
$f(m,m_t,m_p) = f(m/m_t,m_p/m_t)$, which in Dohnanyi's case
was stated
simply as $m_0=Cm_t$ with $C$ constant, and 
$m_B=Bm_t$ with $B$ constant. From Eq. (\ref{eq:fcombo}), we
see that this is equivalent to assuming $m_{0,B} \propto m_t$ and
$Q_D^\star$ is constant. Changing the variables of integration to
$x = z = m/m_t$ and $y = m_p/m_t$ in Eq. (\ref{eq:flux5}), and using
Eq. (\ref{fsplit}) we have
\ba
F(m)\propto m^{11/3-2\alpha}
\int\limits_0^1 dx x^{2\alpha-14/3} \int\limits_0^\infty dy y^{-\alpha} f(x,y).
\label{eq:flux_doh}
\ea
Taking $\alpha=11/6$
results in $F(m)$ being constant in agreement with \citet{Dohnanyi} and 
\citet{Tanaka}, and we discuss the conditions under which the
integrals in Eq. (\ref{eq:flux_doh}) converge in Appendix
\ref{subsect:converge}. We will subsequently call a fragmentation
model with $m_{0,B}=Cm_t$ and $m_B=Bm_t$ a ``Dohnanyi model''. 


\subsection{O'Brien \& Greenberg (2003) and Kobayashi \& Tanaka (2010) case}
\label{subsect:OBG}

\citet{OBG} and \citet{KobayashiTanaka} went one step further and 
considered a power law dependence of the strength as given by
Eq. (\ref{eq:QDstar}). If $s=0$ in Eq. (\ref{eq:QDstar}) (i.e. $\beta
= 1$ in Eq. (\ref{eq:mB})), then this reduces to the Dohnanyi
model. At the same time, \citet{OBG} and \citet{KobayashiTanaka} still took
$m_{0,B}=C m_t$, so in their case $x = z = m/m_t$ and $y =
m_p/m_B(m_t)$. Changing variables again to $x$ and $y$ in
Eq. (\ref{eq:flux5}) and using Eq. (\ref{fsplit}), we find
\ba
F(m)\propto m^{5/3 + (1+\beta)(1-\alpha)}
\int\limits_0^1 dx x^{-8/3 - (1+\beta)(1-\alpha)} \int\limits_0^\infty
dy y^{-\alpha} f(x,y).
\label{eq:flux_obg}
\ea
The mass flux is independent of mass if
\ba
\alpha = \frac{\beta + 8/3}{\beta + 1},
\label{eq:OBG}
\ea
which was derived by \citet{OBG} and
\citet{KobayashiTanaka}, and the reader is again referred to Appendix
\ref{subsect:converge} for the conditions under which the integrals in Eq.
(\ref{eq:flux_obg}) converge. The arguments of \citet{PanSari} are
analogous to 
the calculations of \citet{OBG} and \citet{KobayashiTanaka}, but their
qualitative nature has rid them of the need to worry about
the scaling of $m_0$ with $m_t$.
We will subsequently call a fragmentation model with $m_{0,B} = Cm_t$
and $\beta\neq 1$ an ``OBG model''. 


\subsection{Failure of the power law solution.}
\label{subsect:failure}

We now demonstrate that the power law solution (\ref{eq:PL}) 
does not in general make the mass flux completely independent 
of $m$ for any $\alpha$, unless $m_{0,B} \propto m_t$ as
in \S\ref{subsect:Dohnanyi},\ref{subsect:OBG}. To make the
calculations tractable, we assume a power
law dependence for $m_B$ in the form given by Eq. (\ref{eq:mB}), and
for $m_{0,B}$ in the form given by Eq. (\ref{eq:m0B}).

Using Eq. (\ref{eq:flux5}), we make the definitions
\ba
F_{rem}(m) &\propto& \int\limits_m^\infty dm_t m_t^{5/3-\alpha}
\int\limits_0^\infty dm_p m_p^{-\alpha} f_{rem}(z,y) \\
F_{ej}(m) &\propto& \int\limits_m^\infty dm_t m_t^{5/3-\alpha}
\int\limits_0^\infty dm_p m_p^{-\alpha} f_{ej}(x,y),
\ea
where $F = F_{rem} + F_{ej}$.
Because $x \neq z$, in contrast to the Dohnanyi and OBG models, we
change variables to $z$ and $y$ for the remnant flux and to
$x$ and $y$ for the ejecta flux. This yields
\ba
F_{rem}(m) &\propto& m^{5/3 + (1+\beta)(1-\alpha)}
\int\limits_0^1 dz z^{-8/3 - (1+\beta)(1-\alpha)} \int\limits_0^\infty
dy y^{-\alpha} f_{rem}(z,y) \\
F_{ej}(m) &\propto& m^{\mu^{-1}(5/3 + (1+\beta)(1-\alpha))}
\int\limits_0^{m/m_{0,B}(m)} dx x^{-1 -\mu^{-1}(5/3 +
  (1+\beta)(1-\alpha))} \int\limits_0^\infty dy y^{-\alpha} f_{ej}(x,y).
\label{eq:flux_npl}
\ea
As before, $\alpha$ is given by Eq. (\ref{eq:OBG}) in order for the
$m$ dependence outside both of the integrals to vanish. Now, however, $m$
appears in the upper limit of integration in the integral over $x$ in
the expression for $F_{ej}$, and
unless $m_{0,B}(m) \propto m$ (i.e. $\mu = 1$),
$F(m)$ is {\it not}
independent of $m$. This is one of the key
conclusions of this work, and in the rest of the paper we will
investigate the non-power law behavior of fragmentation 
cascades in detail. 

To keep things simple, we will assume that the
mass flux from remnants is negligible so that $F(m) = F_{ej}(m)$. This
is consistent with Fig. 3 of \citet{KobayashiTanaka}, which shows that
$F_{ej}/F_{rem} \sim 10$ when $Q_D^\star$ is constant.


\section{Non-Power Law Behavior}
\label{nonpl}


From Eq. (\ref{eq:flux_npl}) we see that the mass
flux corresponding
to the power law solution of the OBG model depends only weakly
(logarithmically) on $m$. This
motivates us to look for a solution of Eq. (\ref{massflux1}) 
%
%
in the form
\ba
n(m) \propto m^{-\alpha} \varphi(m), ~~~\alpha = \frac{\beta +
  8/3}{\beta + 1}
\label{eq:nonpl}
\ea
where $\varphi(m)$ is a slowly varying function of $m$:
\ba
\left| \frac{d\varphi}{dm} \right| \ll \left| \frac{\varphi}{m} \right|.
\label{eq:weak_var}
\ea
This property can be verified a posteriori, after the explicit 
form of $\varphi(m)$ is obtained.

With $n(m)$ given by Eq. (\ref{eq:nonpl}), Eq. (\ref{eq:flux5}) becomes
\ba
F(m)\propto \int\limits_m^\infty dm_t m_t^{5/3-\alpha}\varphi(m_t)
\int\limits_{0}^{\infty} 
dm_p m_p^{-\alpha}\varphi(m_p) f_{ej}(x,y).
\label{eq:flux8}
\ea
As discussed in Appendix \ref{subsect:converge}, the value of $y^{-\alpha}
f_{ej}(x,y)$
typically drops off below $y \lesssim k$ and above $y \gtrsim k$, which
means that $m_p^{-\alpha} f_{ej}(x,y)$ is peaked at $m_p \sim
k m_B(m_t)$, where $k$ is a constant. In the case when erosion is
neglected  $k \sim 1$, since collisions with projectiles of mass $m_p <
m_B$ contribute no mass flux, but if erosion is included, then $k \ll
1$  (Fig. 6 of \citet{KobayashiTanaka}). Together with
the condition that $\varphi(m)$ is a slowly
varying function of $m$ (Eq. (\ref{eq:weak_var})), this allows us to expand
$\varphi(m_p)$ in a Taylor series about $m_p = km_B(m_t)$:
\ba
\varphi(m_p) \approx \varphi(km_B(m_t)) + \frac{d \varphi}{d \ln
  m_p}\Big |_{m_p = km_B(m_t)}
\ln \left(\frac{m_p}{km_B(m_t)}\right).
\ea
With this approximation, the inner integral of Eq. (\ref{eq:flux8}) becomes
\ba
\varphi(km_B(m_t))\int\limits_{0}^{\infty} dm_p m_p^{-\alpha}\varphi(m_p)
f_{ej}(x,y)\left[1 + \frac{d \ln \varphi}{d \ln m_p}\Big |_{m_p = km_B(m_t)}
\ln\left(\frac{m_p}{km_B(m_t)}\right)\right].
\ea
From Eq. (\ref{eq:weak_var}), $|d \ln \varphi/d \ln m| \ll 1$, so as long
as the peak in $m_p^{-\alpha} f_{ej}(x,y)$ at $m_p = km_B(m_t)$ is sharp
enough that $\ln(m_p/km_B(m_t)) \sim 1$ over the width of the peak,
then the second term is negligible in comparison with the
first. Thus, up to terms of order $d \ln \varphi/d
\ln m$, Eq. (\ref{eq:flux8}) becomes 
\ba
F(m)\propto \int\limits_m^\infty dm_t m_t^{5/3-\alpha}\varphi(m_t)
\varphi(km_B(m_t)) \int\limits_{0}^{\infty} dm_p m_p^{-\alpha} f_{ej}(x,y).
\label{eq:flux9}
\ea
Changing the inner variable of integration from $m_p$ to $y$,
using the value of $\alpha$ from Eq. (\ref{eq:nonpl}), and using the
definition of $m_B(m_t)$ from Eq. (\ref{eq:mB}), we have
\ba
F(m)\propto \int\limits_m^\infty dm_t m_t^{-1}\varphi(m_t)
\varphi(km_B(m_t)) \int\limits_{0}^{\infty} dy y^{-\alpha} f_{ej}(x,y).
\label{eq:flux10}
\ea

Finally, defining
\ba
f_{ej}(x) \equiv \int_0^\infty dy y^{-\alpha} f_{ej}(x,y),
\label{eq:fejx}
\ea
and introducing the auxiliary function 
\ba
\Theta(m_t)\equiv \varphi(m_t)\varphi(km_B(m_t)).
\label{eq:Theta}
\ea
we arrive at
\ba
F(m)\propto \int\limits_m^\infty dm_t m_t^{-1}\Theta(m_t) f_{ej}(x)
\label{eq:master}
\ea
with $x$ given by Eq. (\ref{eq:x}).
Equation (\ref{eq:master}) is the master equation 
for the two-step determination of $\varphi(m)$: 
\begin{itemize}
\item First, given the explicit form of $f_{ej}(x)$ one needs 
to solve this integral equation under the assumption 
$F=$const to determine the behavior of the auxiliary function 
$\Theta(m)$.
\item Second, having obtained $\Theta(m)$ one must 
solve the functional equation, Eq. (\ref{eq:Theta}), to determine 
$\varphi(m)$. 
\end{itemize}
We now perform this procedure explicitly for two specific forms of $f_{ej}(x)$
--- monodisperse (\S\ref{mononpl}) and power law (\S\ref{sec:gpl}).


\subsection{Monodisperse Fragment Mass Distribution}
\label{mononpl}

We first consider the special case of the {\it monodisperse} 
fragment mass distribution, which puts all fragments at a 
single mass scale $m_0(m_t,m_p)=m_{0,B}(m_t)$:
\ba
g_{ej}(m_f,m_t)=\frac{m_t}{m_{0,B}(m_t)}\delta\left(m_f-m_{0,B}(m_t)\right).
\label{eq:g_md}
\ea
This singular fragmentation model can be thought of as a very crude 
qualitative approximation to any fragmentation law that has 
most of the debris mass concentrated at one scale.
It allows us to obtain some interesting 
analytical results and 
serves as a simple stepping stone for the more general case
considered in \S\ref{sec:gpl}.

The fragmentation law (\ref{eq:g_md}) implies
\begin{equation}
\label{monog}
f_{ej}(x) = \begin{cases} 0, & x < 1 \\
  1, & x > 1 \end{cases},
\end{equation}
where $x = m/m_{0,B}(m_t)$ (\S\ref{fragmentationmodel}).
Plugging this into the master equation (Eq. (\ref{eq:master})) one obtains
\ba
&& F(m)\propto \int\limits_m^{\tilde m_{0,B}(m)} \frac{dm_t}{m_t} 
\Theta(m_t),
\label{eq:flx_monodisp}
\ea
where $\tilde m_{0,B}(m)$ is a new function defined as a
function inverse $m_{0,B}(m)$ (i.e. $m_{0,B}(\tilde m_{0,B}(m)) = m$).
Upon differentiation with respect to $m$, expression 
(\ref{eq:flx_monodisp}) results in 
\ba
\frac{\Theta(m)}{m}=\frac{\Theta\left(\tilde m_{0,B}(m)\right)}
{\tilde m_{0,B}(m)}\frac{d\tilde m_{0,B}(m)}{dm},
\label{eq:monodisp_ansatz}
\ea
where we have used the fact that $F(m)$ is constant.
This functional equation is valid for arbitrary  
$m_{0,B}(m_t)$ provided that the fragmentation law is monodisperse.

We now focus on $\tilde m_{0,B}(m)$ in the form 
\ba
\tilde{m}_{0,B}(m) = \left( \frac{m}{C} \right)^{1/\mu},
\ea
which is valid for $m_{0,B}$ given by Eq. (\ref{eq:m0B}).
Plugging this expression into (\ref{eq:monodisp_ansatz}) we 
find
\ba
\Theta(m)=\frac{1}{\mu}
\Theta\left(\left(\frac{m}{C}\right)^{1/\mu}\right).
\label{eq:monodisp_pl}
\ea

Introducing the new variable 
$t \equiv \ln(m^{1-\mu}/C) = \ln(m/m_{0,B}(m))$ 
and the new function\footnote{We are grateful to Jeremy Goodman for 
suggesting this transformation.} 
$\Theta_1(t) \equiv \Theta\left(\left(Ce^{t}\right)^{1/(1-\mu)}\right)$, 
Eq. (\ref{eq:monodisp_pl}) becomes
\begin{equation}
\label{psi1eq}
\Theta_1(t) = \frac{1}{\mu}\Theta_1\left(\frac{t}{\mu}\right).
\end{equation}
This has the form of a homogeneous functional equation 
\begin{equation} 
\label{funceq}
q(au) = bq(u),
\end{equation}
($a$ and $b$ are constants) which has the solution 
\begin{equation}
q(u) = T(\ln u)u^\lambda, \ \ \lambda \equiv \frac{\ln b}{\ln a}.
\end{equation}
Here $T(s) = T(s + \ln a)$ is an arbitrary periodic function with
period $\ln a$, which can be constant \citep{Polyanin}. This implies
that the solution of Eq. (\ref{psi1eq}) is
\begin{equation}
\Theta_1(t) = \frac{T(\ln t)}{t},
\end{equation}
so that finally
\begin{equation}
\label{psisol}
\Theta(m) = \frac{T(\ln(\ln(m/m_{0,B}(m)))}{\ln(m/m_{0,B}(m))}, \ \ T(s) = T(s
+ \ln \mu). 
\end{equation}

Clearly this solution is inapplicable to the OBG case of $\mu = 1$,
because the variable $t$ then reduces to a constant. However, for
$\mu = 1$, Eq. (\ref{eq:monodisp_pl}) already has the form 
(\ref{funceq}), so that its solution is
\begin{equation}
\Theta(m) = T(\ln m), \ \ T(s) = T(s + \ln C).
\end{equation}
In particular, $\Theta(m)$=const, and consequently  
$\varphi(m)$=const, is one of the possible solutions, 
so that $n(m)$ simply proportional to $m^{-\alpha}$ with 
$\alpha$ given by (\ref{eq:OBG}) is a viable solution for a 
fragmentation cascade with $\mu=1$, in agreement with \citet{OBG}. 

However, the existence of periodic solutions brings about
the possibility of having {\it waves} in the mass distribution 
of objects while still having $F(m)$ constant, even for $\mu = 1$. 
The presence of waves at masses $m/m_{rm} \sim 1$ in fragmentation
cascades having a lower mass
cutoff has been previously demonstrated by
\citet{CampoBagatin}, \citet{DD}, and \citet{Thebault};
\citet{OBG} found waves to appear whenever
the scaling of specific energy necessary for disruption $Q_D^\star$ 
with object mass changed abruptly (e.g. due to an object's 
self-gravity becoming more important than its internal strength); and
\citet{Fraser}, \citet{PanSari}, and \citet{KB2004} have shown waves
to be present at the transition from a collisionally
evolved to a primordial size distribution (i.e. at $m/m_{inj} \sim 1$).

The nature of the waves we have found is different from the 
ones discussed by previous authors, since they exist even when
$m_{rm} = 0$, $m_{inj} = \infty$, and $m_B$ 
is given by a pure power law without breaks (Eq. (\ref{eq:mB})). In
astrophysical systems, these kinds of waves could be
triggered in stochastic
collisions of large planetesimals \citep{KB2005, WyattDent,
  Wyatt}. Most of the mass in such a collision would be in particles
of size $m_0(m_t,m_p)$, and if the density of particles with mass $m_0$ created
in the collision is comparable to or exceeds the local disk density of
such particles, a wave will be triggered. However,
a proper treatment of these
waves needs to account for a non-monodisperse fragment mass 
distribution, which could damp them, so we leave this subject 
for future work (Belyaev \& Rafikov, in preparation).


\subsection{Power Law Fragment Mass Distribution}
\label{sec:gpl}

We now assume that the mass spectrum of fragments produced in 
a collision $g_{ej}(m_f,m_t,m_p)$
is a power law with an index $-\eta$ having 
a cutoff at a maximum fragment mass $m_0(m_t,m_p) = m_{0,B}(m_t)$. This
allows us to write  
\begin{eqnarray}
\label{Geq}
f_{ej}(x) =  \begin{cases} x^{2-\eta}, & x < 1 \\
  1, & x > 1 \end{cases},
\end{eqnarray}
where as before $x = m/m_{0,B}(m_t)$ (\S\ref{fragmentationmodel}).
This fragmentation model resembles reality, since observational 
and experimental evidence \citep{Gault,Fujiwara} as well as numerical 
simulations \citep{BAR,BenzAsphaug} suggest power 
law behavior of $g_{ej}(m,m_t,m_p)$ 
at small fragment masses (\S\ref{fragmentationmodel}). Various flavors
of such a power law model have been adopted in 
theoretical studies by \citet{Dohnanyi}, \citet{Williams},
\citet{OBG}, \citet{KB},
\citet{DavisFarinella}, \citet{KobayashiTanaka}, etc. 

Plugging Eq. (\ref{Geq}) into the master equation (Eq. 
(\ref{eq:master})), we find
\ba
&& F(m)\propto \int\limits_m^{\tilde m_{0,B}(m)} \frac{dm_t}{m_t} 
\Theta(m_t)+ \int\limits_{\tilde m_{0,B}(m)}^\infty 
\frac{dm_t}{m_t}\left[\frac{m}{m_{0,B}(m_t)}\right]^{2-\eta} 
\Theta(m_t).
\label{eq:flx_PL}
\ea
Differentiating this expression with respect to $m$  
(keeping in mind that $dF(m)/dm=0$), dividing the resultant expression 
by $m^{1-\eta}$, and differentiating again one finds
\ba
\frac{d\ln\tilde m_{0,B}(m)}{d\ln m}\Theta\left(\tilde m_{0,B}(m)\right)=
\Theta(m)-\frac{1}{2-\eta}\frac{d\Theta(m)}{d\ln m}.
\label{eq:PL_ansatz}
\ea
This equation reduces to Eq. (\ref{eq:monodisp_ansatz}) if one
takes the limit $\eta\to-\infty$ which corresponds to all of the fragments'
mass concentrated in objects with mass $m_{0,B}$, and is thus 
equivalent to a monodisperse fragmentation law.

Until now, our treatment was rather general and based solely 
on assumption (\ref{Geq}) so that the functional equation 
(\ref{eq:PL_ansatz}) is valid for any form of $m_{0,B}$.
We now take $m_{0,B}$ in the form of Eq. (\ref{eq:m0B}) and find
\begin{equation}
\frac{\Theta(\tilde m_{0,B}(m))}{\mu} = \Theta(m) - \frac{1}{2-\eta}
\frac{d\Theta(m)}{d\ln m}.
\label{diffeqpowlaw}
\end{equation}
In the limit $\eta \to -\infty$, corresponding to the
monodisperse case, the second term on the right hand side of
(\ref{diffeqpowlaw}) is zero, and the solution is easily verified to be 
\begin{equation}
\Theta(m) = \frac{1}{\ln(m/m_{0,B}(m))}.
\label{thetamonodisperse}
\end{equation}
This solution could also have been obtained from Eq. (\ref{psisol})
by setting $T=1$.

For the non-monodisperse case, we can solve Eq.
(\ref{diffeqpowlaw}) by first introducing a new independent
variable $w \equiv 1/\ln(m/m_{0,B}(m))$, and 
the new function $\Theta_1(w)=\Theta(m)=\Theta((Ce^{1/w})^{1/(1-\mu)})$. 
This is a natural choice, since $\Theta(m)=w$ is 
the exact solution for the monodisperse case if 
$T(\ln(\ln(m/m_{0,B}(m))) = 1$.
With these definitions we can rewrite Eq. 
(\ref{diffeqpowlaw}) as
\begin{eqnarray}
\frac{\Theta_1(\mu w)}{\mu}=\Theta_1(w)-
\frac{\mu-1}{2-\eta}w^2\frac{d\Theta_1(w)}{dw}.
\label{eq:fundif}
\end{eqnarray} 
We next look for the solution of this equation in the form
of an infinite series
\ba
\Theta_1(w) = \sum\limits_{k=1}^\infty A_k w^k
\label{eq:series}
\ea
(note that $A_0=0$ as long as $\mu\neq 1$). By plugging this ansatz
into Eq. 
(\ref{eq:fundif}) and changing the index of summation in the last 
term we get 
\begin{eqnarray}
\label{psiexact}
\sum_{k=1}^\infty A_k \mu^{k-1} w^k =
\sum_{k=1}^\infty A_k w^k + \frac{1-\mu}{2-\eta}\sum_{k=2}^\infty
(k-1) A_{k-1} w^{k}.
\end{eqnarray}
We can set $A_1$ equal to any value and this corresponds 
to an overall normalization of $\Theta_1$. We then obtain 
the other coefficients from the recursive relation
\begin{equation}
A_{k+1} = \frac{(1-\mu)k}{(2-\eta)(\mu^k - 1)} A_k,~~~k=1,..,\infty.
\label{serieseq}
\end{equation}
Unfortunately, this series only converges for $\mu > 1$. 

For $\mu
< 1$, we have found numerically in \S \ref{sec:mbm_pow} that the formula 
\begin{equation}
\label{psiapprox}
\Theta(m) = \frac{1}{\ln(m/m_{0,B}(m)) + 1/(2-\eta)}
\end{equation}
gives good results up to $\eta \lesssim 1.7$. Equation
(\ref{psiapprox}) is a version of the monodisperse solution
(\ref{thetamonodisperse}), which has
been shifted in $\ln m$, and is accurate up to terms of
$O(w^3)$ in Eq. (\ref{psiexact}).


\section{Numerical verification of non-power law behavior}
\label{numerical}


Having obtained solutions for $\Theta(m)$ for a couple of  
fragmentation models, we are now in a position to
determine $\varphi(m)$ from Eq. 
(\ref{eq:Theta}). The analytical calculations involved 
in this process are rather cumbersome and we refer 
the interested reader to Appendix 
\ref{app:varphi} for the mathematical details. 
There, we describe the general method of 
solving for $\varphi(m)$ which works 
%
%
for $m_B$ given by Eq. (\ref{eq:mB}) and for 
arbitrary $\Theta(m)$.
%
%
However, we provide explicit analytical results only for 
$\Theta(m)$ corresponding to the 
monodisperse case in Appendix \ref{app:monodisp}. In this section, we
compare these analytical results with
numerical calculations of fragmentation cascades. The latter
were carried out using a fragmentation code which is described 
in detail in Appendix \ref{codeappendix}. For simplicity, we ignore
erosion in our calculations, which amounts to setting $k=1$ in
Eq. (\ref{eq:Theta}).


\subsection{Results for $\mu\neq 1$ and $m_B(m) = m$.}
\label{subsect:mB=m}

As discussed in \S\ref{simplifications}, real astrophysical systems
typically have $m_B(m_t) \ll m_t$. In order to {\it qualitatively}
understand the non-power law behavior, however, it is instructive to
consider a simplified model in which there is no erosion. Next, we
also assume that a target can only be broken by
projectiles that are of very nearly the same size as itself: 
\ba
m_B(m_t)=(1-\epsilon) m_t,~~~\epsilon\ll 1.
\label{eq:mB=m}
\ea
Although such an assumption is unrealistic, it is useful for getting a
qualitative picture of the non-power law behavior. 

In the
model we have just introduced, which we will call the $m_B = m$ model,
the analog of Eq. (\ref{eq:flux8}) is
\ba
F(m)\propto \int\limits_m^\infty dm_t m_t^{5/3-\alpha}\varphi(m_t)
\int\limits_{m_B(m_t)}^{\tilde{m}_B(m_t)} 
dm_p m_p^{-\alpha}\varphi(m_p) f_{ej}(x,y),
\label{eq:fluxmB=m}
\ea
where $\tilde{m_B}(m)$ is defined to be the largest
mass that can be broken by a projectile of mass $m$
(i.e. $m_B(\tilde{m_B}(m)) = m$). The power law
slope of this model is $\alpha = 11/6$ just like the Dohnanyi case,
and it is straightforward to show that the master
equation (Eq. (\ref{eq:master})) and Eq. (\ref{eq:Theta}) are both still
valid for the $m_B = m$ model. Now, however, Eq. (\ref{eq:Theta}) is
trivial to solve, since we have
$m_B(m_t) \approx m_t$, which implies that $\varphi(m) \approx
\sqrt{\Theta(m)}$. Comparison between the
numerical and analytical behavior of $\varphi(m)$ then becomes 
a direct verification of our solutions for $\Theta(m)$ obtained 
in \S\ref{nonpl}, and we avoid the
intermediate steps involved in solving  Eq. 
(\ref{eq:Theta}) for arbitrary $m_B$.

In our numerical calculations we implement the case of 
$m_B(m_t) = m_t$  by taking $m_B(m_t)=0.99 m_t$. We specify the
amount of time for which
simulation was run in one of two ways, but both make use of the
collision time defined generally as 
\begin{equation}
\label{tauc}
\tau_c(m_t,t)  \equiv \left[ \left( \frac{3}{4 \pi \rho} \right)^{2/3} 
\pi v \int\limits_{m_B(m_t)}^{\tilde m_B(m_t)} dm_p n(m_p,t)
\left(m_p^{1/3}+m_t^{1/3}\right)^2 \right]^{-1}. 
\end{equation}
Here, we have assumed a geometric cross-section for collisions and
mass-independent relative velocities as in \S
\ref{simplifications}. First, we can specify how long a simulation was
run in terms of the number
of collision times of the smallest particles in the simulation. This
is a natural unit of measure to use, since this is the
timescale on which the low mass end of the particle mass distribution
evolves, unless the initial distribution is already in a
steady-state. Second, we can specify how long a simulation was run by the
location of the collisional break, $m_{break}$, at the end of the
simulation, $t = t_{end}$, which is given
implicitly by $\tau_c(m_{break},t_{end}) = t_{end}$. The quantity
$m_{break}$ is especially useful when interpreting simulation results
for which the distribution was initialized to be in a steady-state to
check an analytical steady-state solution (\S \ref{plfsd} and
\ref{subsect:mBgen}). In this
case, there should be no evolution of the distribution in time, and so no break
should develop. However, $m_{break}$ gives the mass below which we {\it
  would have seen} evolution, if the analytical steady-state solution
were incorrect. 

\subsubsection{Monodisperse Fragment Mass Distribution.}
\label{sec:mbm_pow}

We first consider the ansatz (\ref{eq:mB=m}) with $\mu\neq 1$
and the monodisperse 
fragment mass distribution. Using Eq. (\ref{psisol}) we find
\begin{equation}
\label{phimbm}
\varphi(m) = \frac{1}{\sqrt{\ln(m/m_{0,B}(m))}}, 
\end{equation}
where we have set the arbitrary periodic function $T$ to a constant.

\begin{figure}[!h]
  \centering
  \subfigure[]{\includegraphics[width=.49\textwidth]{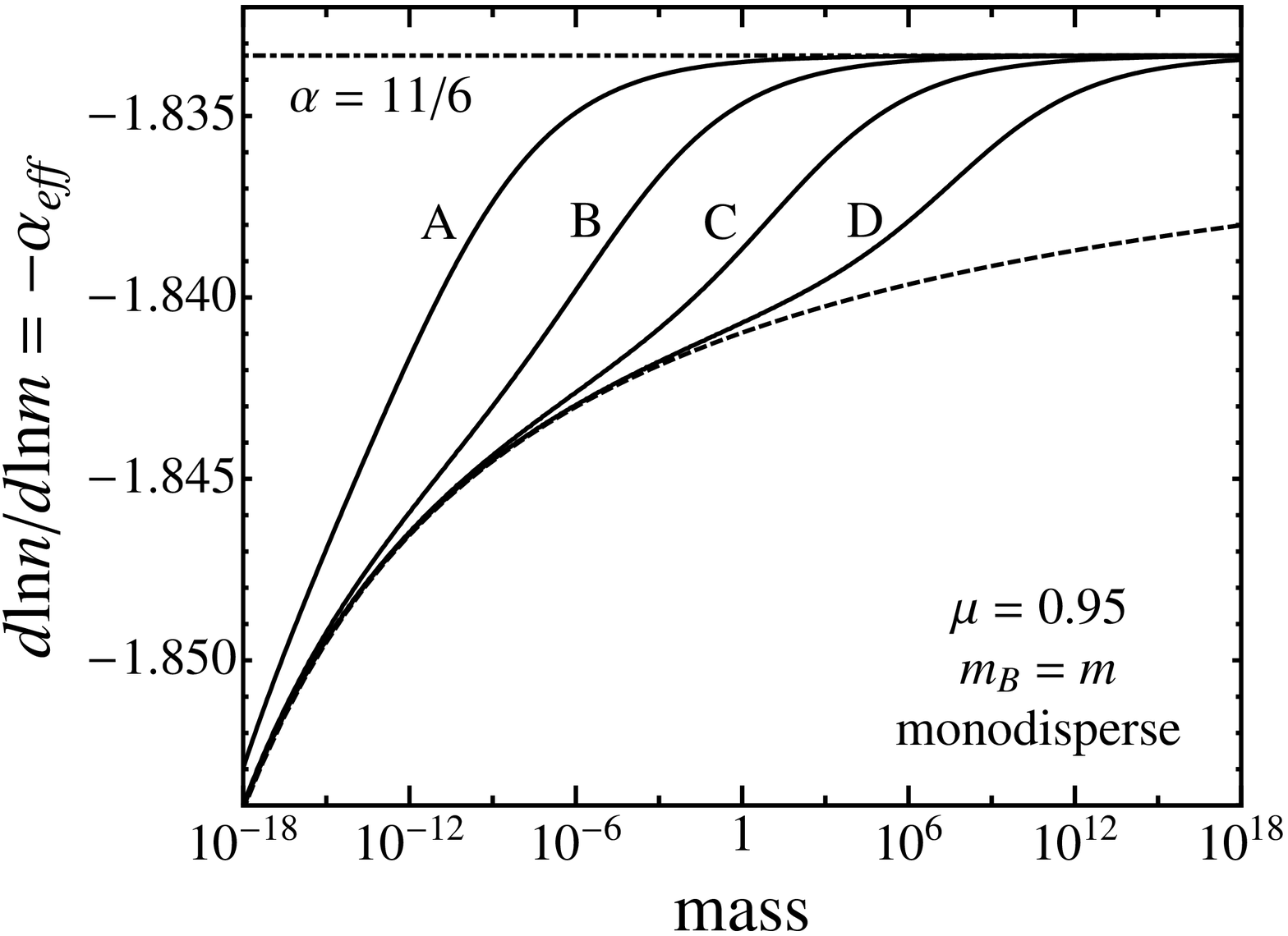}}
  \subfigure[]{\includegraphics[width=.49\textwidth]{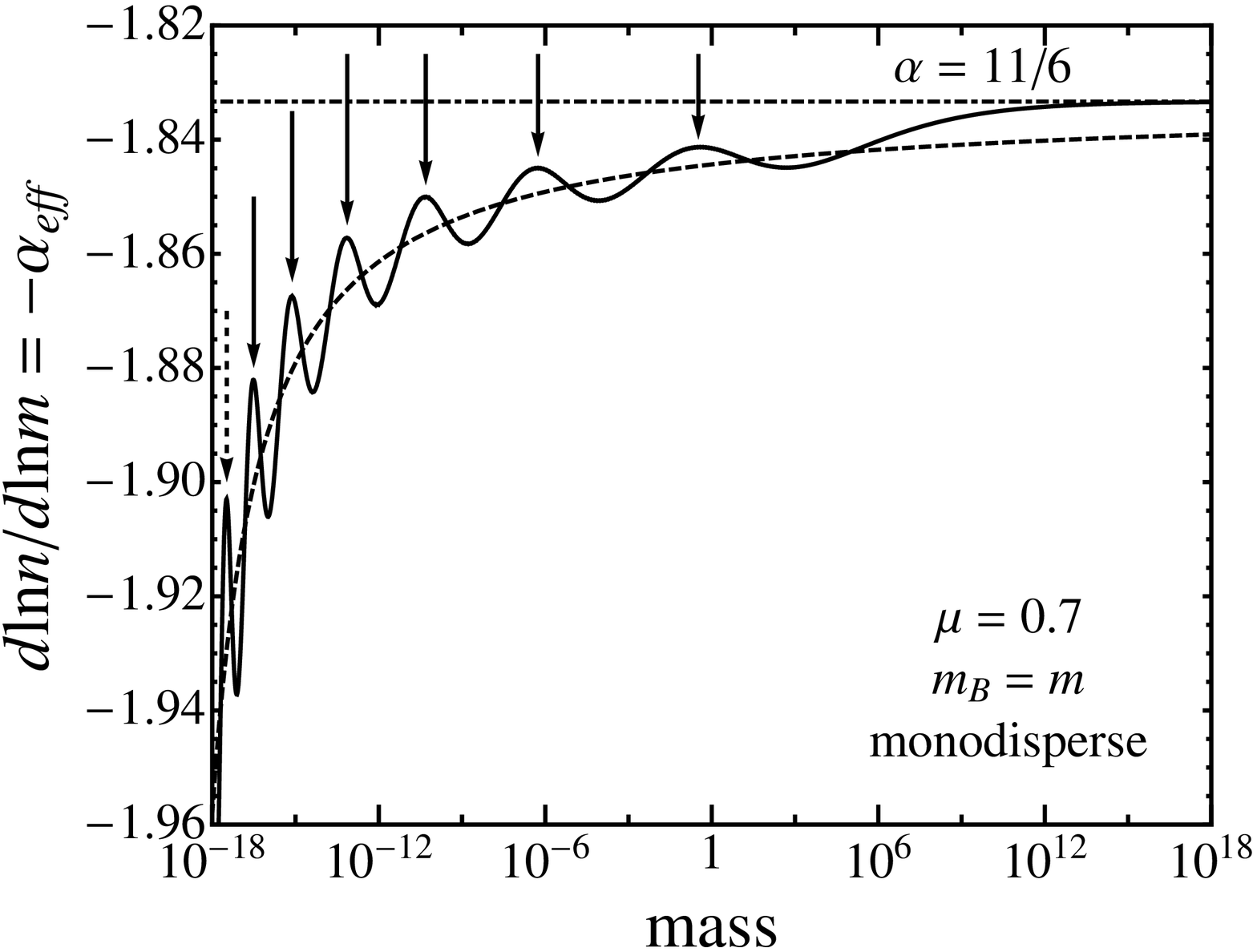}} 
  \caption{Population index $d\ln n/d\ln m = -\alpha_{eff}$ of the particle 
mass distribution vs. mass for two different values of 
$\mu$ in the relation $m_{0,B}=Cm_t^\mu$. Both panels display the evolution 
of $\alpha_{eff}$ for $m_B(m) = m$ starting from an initial 
distribution $n_0(m)\propto m^{-11/6}$ (the Dohnanyi model). 
Panel (a) 
shows the smooth convergence of $n(m,t)$ from $n_0(m)$ 
to the solution $n(m) = m^{-\alpha} \varphi(m)$, with
$\varphi(m)$ given by Eq. (\ref{phimbm}) and shown by 
the dashed line. The model parameters are $m_B(m) = m$, 
$\mu = 0.95$, and $m_{0,B}(m=10^{-18}) = 0.3m$ (mass scale is arbitrary), 
which sets $C$ in the expression for $m_{0,B}$. The curves A-D show the
solution after 25, 250, 2500, and 25000 collision times (defined in \S \ref{subsect:mB=m})
of particles with mass $m =
10^{-18}$. 
Panel (b)  shows the numerical solution starting again from a Dohnanyi
distribution as in panel (a), but with $\mu=0.7$ and all other parameters
the same. The simulation was run for 5000 collision times of particles
with mass $m = 10^{-18}$. The numerical solution 
(solid line) no longer converges smoothly to the analytic solution 
(\ref{phimbm}) (dashed line), but oscillates about it. 
Given the location of the leftmost peak (dashed arrow), Eq.
(\ref{psisol}) correctly predicts the locations of the other 
peaks as indicated by the solid arrows.} 
 \label{mbmfig}
\end{figure}

To verify Eq. (\ref{phimbm}) numerically, we initialize
a pure power law mass distribution 
$n_0(m)\propto m^{-\alpha}$ with an index $\alpha = 11/6$,
which is the steady-state solution of the Dohnanyi model. 
We then expect that for $\mu\neq 1$ the shape of $n(m,t)$ will
gradually evolve from $n_0(m)$ towards the correct solution 
(\ref{eq:nonpl}) with $\varphi(m)$ given by Eq. 
(\ref{phimbm}). To illustrate this evolution in Fig. \ref{mbmfig} 
and all subsequent figures, we plot the ``effective population index'' 
$\alpha_{eff}(m,t)\equiv - d\ln n(m,t)/d\ln m$ 
as a function of $m$ at different moments of time. This way of 
representing the evolution of $n(m,t)$ naturally highlights the 
non-power law behavior, since $\alpha_{eff}$ for $n$ given by a
power law as in Eq. (\ref{eq:PL}) appears as the 
horizontal line $\alpha_{eff}=\alpha$ in such plot. Thus, any 
deviation from a horizontal line is indicative of non-power 
law scaling of $n(m)$. 

In Fig. \ref{mbmfig}a, we display the case of $\mu$ close 
to unity ($\mu = 0.95$), which according to (\ref{phimbm}) 
corresponds to an almost constant $\varphi$ since  
$m/m_{0,B}(m)\propto m^{0.05}$; note the small 
range of variation of $\alpha_{eff}$ in Fig. \ref{mbmfig}a. 
For this value of $\mu$ we indeed find that the initial power law 
distribution smoothly converges to the analytic solution 
(\ref{phimbm}) over time. 

In Fig. \ref{mbmfig}b, we show the evolution of $\alpha_{eff}$
for $\mu = 0.7$, appreciably different from $\mu=1$. One
can see that in this case the numerical solution does not 
converge towards the analytical solution in a smooth fashion. 
Instead the numerical solution {\it oscillates} 
in mass space about the analytical solution. 
The frequency of these oscillations is correctly predicted by 
Eq. (\ref{psisol}), and the
positions of the peaks computed according to this formula 
are shown by the arrows in Fig. \ref{mbmfig}b. 
Their agreement with the locations of numerical peaks proves 
that waves can indeed be spontaneously generated and persist with 
no indication of damping in smooth fragmentation models (i.e. for 
functions $m_B$ and $m_0$ not having any breaks caused by the 
abrupt changes of the material properties of colliding objects), 
at least for a monodisperse fragment mass distribution.

In this study, the appearance of waves is undesirable as it
complicates the comparison between numerical and analytical 
solutions for $n(m)$, and one would like to avoid their 
excitation. Comparing the two cases depicted in Fig. 
\ref{mbmfig} suggests that waves get generated when the initial 
distribution $n_0(m)$ is significantly different from the
analytical, non-oscillatory steady-state solution. Motivated 
by this observation, we start from 
an initial distribution $n_0(m)$ which is {\it identical} 
to the analytical steady-state 
solution in subsequent calculations. We then expect that the
numerical solution, $n(m,t)$, will not deviate 
from $n_0(m)$ as time goes by if our steady-state solution is 
correct; if it is not, then $n(m,t)$ will evolve away from $n_0(m)$.

Despite the complications related to the appearance of waves, 
it is clear that in the monodisperse case, $n(m)$
exhibits non-power law behavior for $\mu\neq 1$, and our
analytical solutions (\ref{psisol}) and (\ref{phimbm}) accurately 
describe the deviation from the pure power law solution.

To better understand the qualitative behavior of the steady-state 
$\alpha_{eff}(m)$ in Fig. \ref{mbmfig}, we use Eq.
(\ref{eq:m0B}), Eq. (\ref{eq:nonpl}), and Eq. (\ref{phimbm}) to write
\ba
\alpha_{eff}=\alpha+\frac{1}{2\ln(m/m_0^\star)},
\label{eq:alpha_eff_monodisp}
\ea
where $m_0^\star = C^{1/(1-\mu)}$ is the mass scale at which the mass 
of the largest fragment becomes formally equal 
to the mass of the target (i.e. $m_{0,B}(m_0^\star) = m_0^\star$). For
$\mu<1$ a fragmentation
cascade can only exist for $m>m_0^\star$, in which case
$\alpha_{eff}>\alpha$, as can be seen in Fig. \ref{mbmfig}
($\alpha=11/6$, since $\beta=1$ for $m_B=m$). Thus,
in the monodisperse case with $m_B = m$, the $\mu<1$ collisional mass 
spectrum is {\it steeper} than in the Dohnanyi model. The deviation
of $\alpha_{eff}$ from the Dohnanyi slope increases as
$m\to m_0^\star$, and decreases for $m\gg m_0^\star$.
For example, in the case shown in Fig. \ref{mbmfig}b
one has $ m_0^\star=1.8\times 10^{-20}$ and $\alpha_{eff}$
deviates from 11/6 by $\approx 0.12$ already at $m=10^{-18}$, 
i.e. at $m/m_0^\star\sim 55$.

\begin{figure}[!h]
  \centering
  \subfigure[]{\includegraphics[width=.49\textwidth]{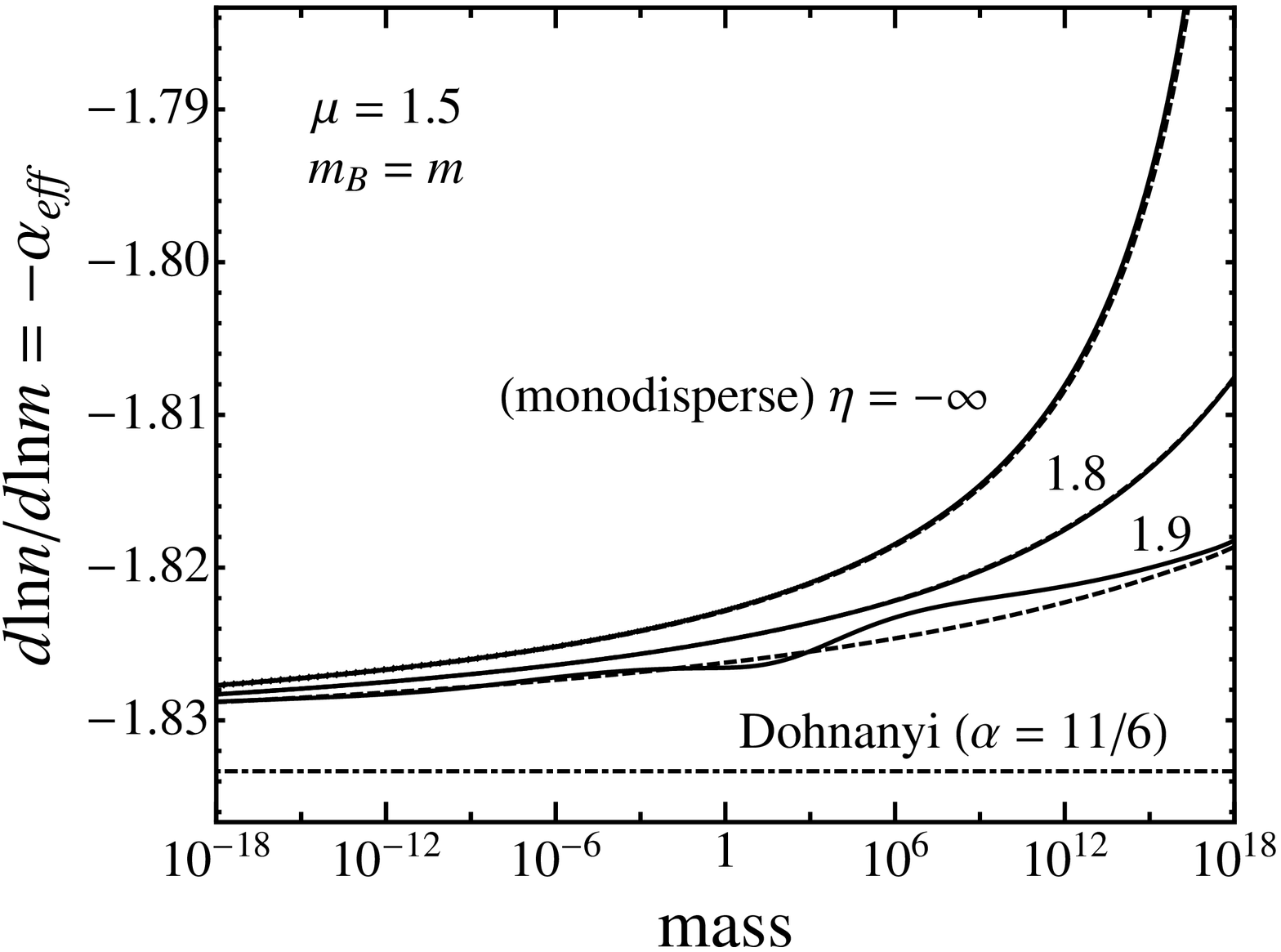}} 
  \subfigure[]{\includegraphics[width=.49\textwidth]{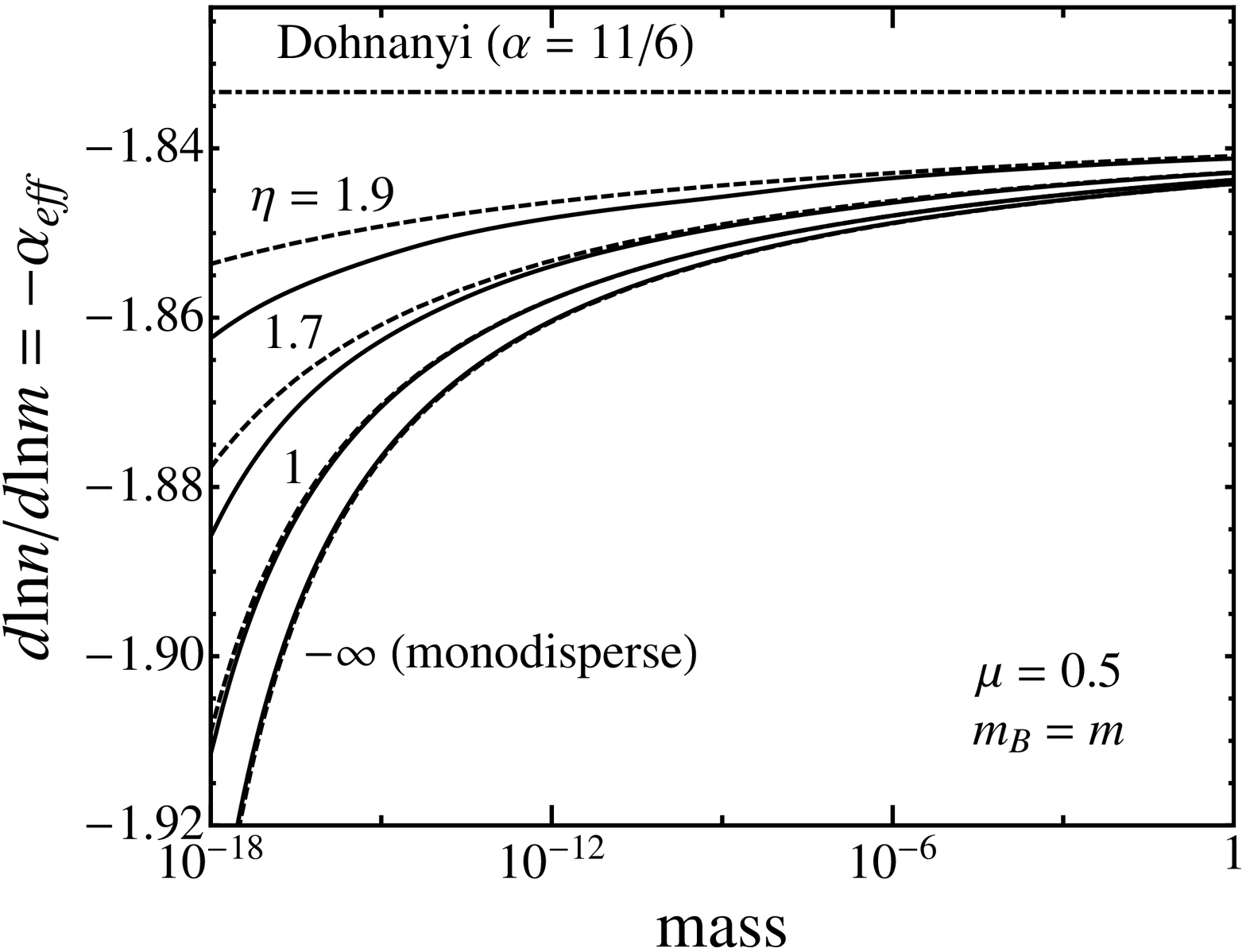}}
  \caption{Population index of the mass distribution
vs. mass for a power law fragment mass distribution and two different 
values of $\mu$. $n_0(m)$ was initialized to be the analytical 
steady-state solution for each model. The dot-dashed line indicates 
the $\alpha = 11/6$ ``Dohnanyi'' power law index which would be 
expected for the $\mu = 1$ case. The labels in each panel 
indicate the value of power law index $\eta$ of the fragment mass 
distribution for different models, and a monodisperse model 
(corresponding to $\eta = -\infty$) is plotted for reference 
(its behavior is given by Eq. (\ref{phimbm})). Panel (a): The
numerical calculations were compared with the analytic series solution 
(Eq. (\ref{psiexact})) truncated after 20 terms. Curves for two
models both having $m_B(m) = m$, $\mu =1.5$, $m_{0,B}(m=10^{18}) =
0.1m$, and either $\eta = 1.8$ or 1.9 are shown. The analytic 
solutions (dashed lines) agree well with numerical runs (solid
lines), although there is some deviation for the $\eta = 1.9$
case. The location of $m_{break}$ (\S \ref{subsect:mB=m}) in the
simulations is $m_{break} \sim 10^3$. Panel (b):
Similar to panel (a), but now the approximate solution
(Eq. (\ref{psiapprox})) was computed for three models, each
having $m_B(m) = m$, $\mu = 0.5$, $m_{0,B}(m=10^{-18}) = 0.1m$, and
either $\eta = 1$, 1.7, or 1.9. There is good agreement between the 
analytic solutions (dashed line) and numerical results (solid lines) 
for $\eta \lesssim 1.7$. 
%
%
Here $m_{break} =
10^5$, although we have truncated the mass range at $m=1$ to highlight
the differences between the solutions.
}
 \label{powlawfig}
\end{figure}

\begin{figure}[!h]
  \centering
  \subfigure[]{\includegraphics[width=.49\textwidth]{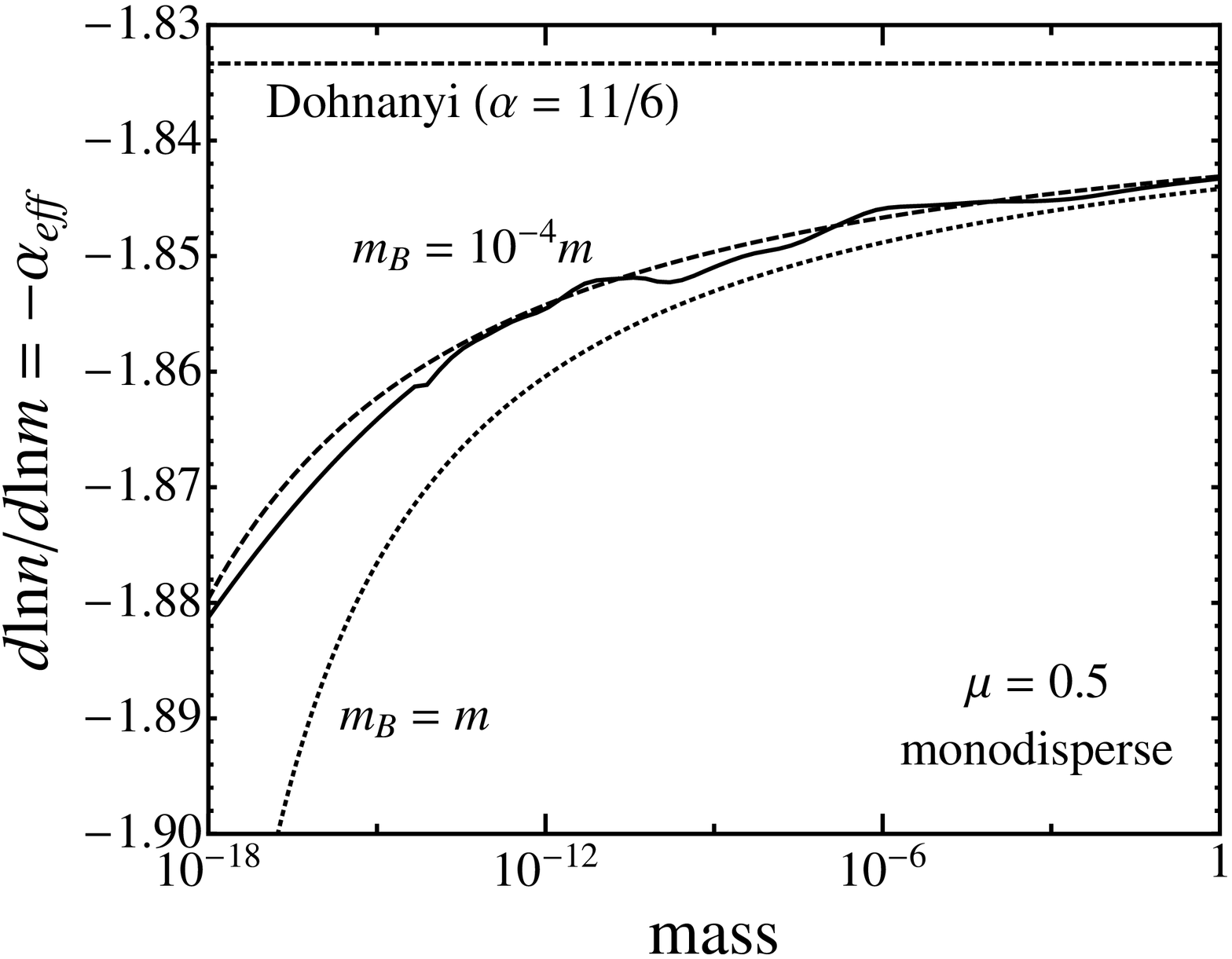}} 
  \subfigure[]{\includegraphics[width=.49\textwidth]{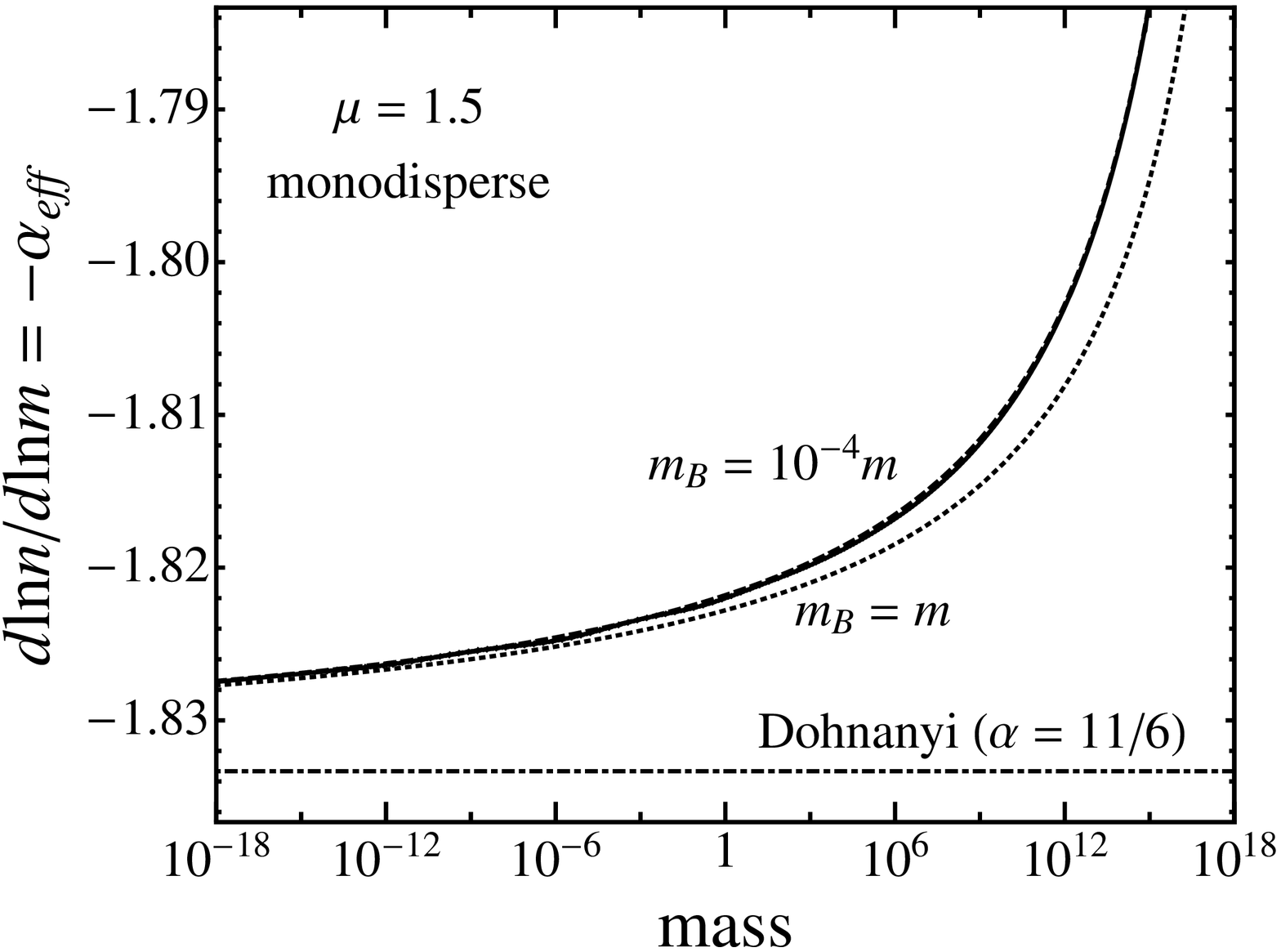}}
  \caption{Population index of the mass distribution
vs. mass for $m_B(m)=Bm$, $B=10^{-4}$, and two different values of
$\mu$. $n_0(m)$ was initialized to be the analytical 
steady-state solution for each model. The dot-dashed line indicates 
the $\alpha = 11/6$ ``Dohnanyi'' power law index which would be 
expected for the $\mu = 1$ case. The dotted line represents 
the monodisperse solution with $m_B(m) = m$.
Panel (a): 
Analytic solution (\ref{eq:varphi_sol_2_norm}),
(dashed line) is compared with the numerical result (solid line) 
for a monodisperse fragmentation 
model having $\mu =0.5$, and $m_{0,B}(m=10^{-18}) = 0.1m$. The location of
$m_{break}$ is at $m_{break} \sim 0.01$. The effect of the imposed 
boundary conditions (Appendix \ref{codeappendix}) is evident as 
the straight line between $m = 10^{-14} - 10^{-18}$. 
Panel (b): 
Similar to panel (a), but now the analytic solution 
(\ref{eq:varphi_sol_1_norm}) was computed for a model with 
$\mu = 1.5$, and $m_{0,B}(m=10^{18}) = 0.1m$. The dashed and solid lines
are hard to distinguish here because the numerical solution 
does not significantly evolve away from the
initial analytical solution. The value of $m_{break}$ is $m_{break} = 10^7$.}
 \label{mbnem_Dohnanyi_fig}
\end{figure}

On the contrary, in the monodisperse case with $\mu>1$,
a fragmentation cascade is possible only for $m<m_0^\star$, and
Eq. (\ref{eq:alpha_eff_monodisp}) then predicts that 
$\alpha_{eff}<\alpha$. Thus, the collisional mass spectrum for 
$\mu>1$ has a {\it shallower} slope than the Dohnanyi solution.
This can be seen in Fig. \ref{powlawfig}a (curve labeled 
``$-\infty$''), Fig. \ref{mbnem_Dohnanyi_fig}b (dotted curve), 
and Fig. \ref{mbnem_nonDohnanyi_fig}a (dotted curve).
In that case, the biggest deviations of $\alpha_{eff}$ 
from $\alpha$ are observed at large masses, as 
$m\to m_0^\star$. 

\subsubsection{Power Law Fragment Mass Distribution}
\label{plfsd}

We now consider ansatz (\ref{eq:mB=m}) with the power law fragment 
mass distribution explored in \S\ref{sec:gpl}. In this case, 
$\varphi(m)$ is given exactly as the square root of Eq. 
(\ref{psiexact}) for $\mu > 1$, and approximately as the square 
root of Eq. (\ref{psiapprox}) for $\mu < 1$. We again 
test these solutions numerically, but this time starting with 
the analytical steady-state mass distribution (as discussed in
the previous section), since we are not interested in waves. 

For the $\mu > 1$ case, we find that the solution given by the
the series in Eq. (\ref{psiexact}) converges quickly, and
truncating the series after the first twenty terms gives a
result which shows little sign of evolution for $\eta =
1.8$, (Fig. \ref{powlawfig}a). Since in the $\eta =1.8$ case
numerical $n(m,t)$
does not evolve significantly from the initial distribution
given by the steady-state solution (Eq. (\ref{psiexact})), this 
implies that our analytic solution for $\mu > 1$ is indeed 
correct, even for $\eta$ very close to 2. 
In Fig. \ref{powlawfig}b we show the analogous calculation for 
the $\mu < 1$ case. We initialize $n_0$ to be the
approximate solution given by Eq. (\ref{psiapprox}) and find that for
$\eta \lesssim 1.7$ this solution works well. However, 
for larger values of $\eta$ deviations between the steady-state
numerical and analytical solutions become
apparent. Nevertheless, the general qualitative behavior is still
reproduced by Eq. (\ref{psiapprox}) even for $\eta$ close to 2. 

It is easy to see from Fig. \ref{powlawfig} that as $\eta \to 2$ the
behavior of $\alpha_{eff}$
flattens out and approaches a power law solution with 
slope given by Eq. (\ref{eq:OBG}), which is just $\alpha=11/6$ 
for $m_B=m$. Such behavior can be understood by looking at Eq. 
(\ref{diffeqpowlaw}), which reduces to $d\Theta(m)/dm = 0$, 
in the limit of $\eta\to 2$. This means that $\Theta(m)$ does 
not vary with mass in this limit, which implies
that $\varphi(m)$ is also constant if $m_B(m) = m$. 
Thus, in the limit $\eta\to 2$ when fragments produced in an individual 
collision are uniformly distributed in $\ln m$, the steady-state
solution is a power law. 

On the other hand, the smaller $\eta$ becomes, the more the fragment mass
distribution $g$ is dominated (by mass) by the largest fragments, which 
makes it approach a monodisperse distribution. One then
expects that when $\eta$ is reduced, $\alpha_{eff}$ should tend
towards the monodisperse form given in
Eq. (\ref{eq:alpha_eff_monodisp}), and this
trend is clearly seen in Fig. \ref{powlawfig}. It is also worth
noting that for a given value of $\mu$, the largest deviation 
of $\alpha_{eff}$ from pure power law behavior occurs 
for the monodisperse fragment mass distribution, which is obtained in
the limit $\eta \rightarrow -\infty$.

\subsection{Results for $\mu\neq 1$ and 
general $m_B(m)$.}
\label{subsect:mBgen}

We now describe our results for the more realistic case when 
$m_B \neq m_t$. $m_{0,B}$ is again given by Eq. (\ref{eq:m0B}) with
$\mu\neq 1$, which is necessary for the non-power law
behavior. The general approach
developed in Appendix \ref{app:varphi} allows us to 
compute $\varphi(m)$ for arbitrary $\Theta(m)$, including 
that obtained in \S\ref{sec:gpl} for the power law fragment 
mass distribution. To keep things simple, however, we only provide
a comparison between numerical
and analytical results in the case of a monodisperse fragment mass 
distribution, for which we derived closed form analytical 
expressions in Appendix \ref{app:varphi}.

\begin{figure}[!h]
  \centering
  \subfigure[]{\includegraphics[width=.49\textwidth]{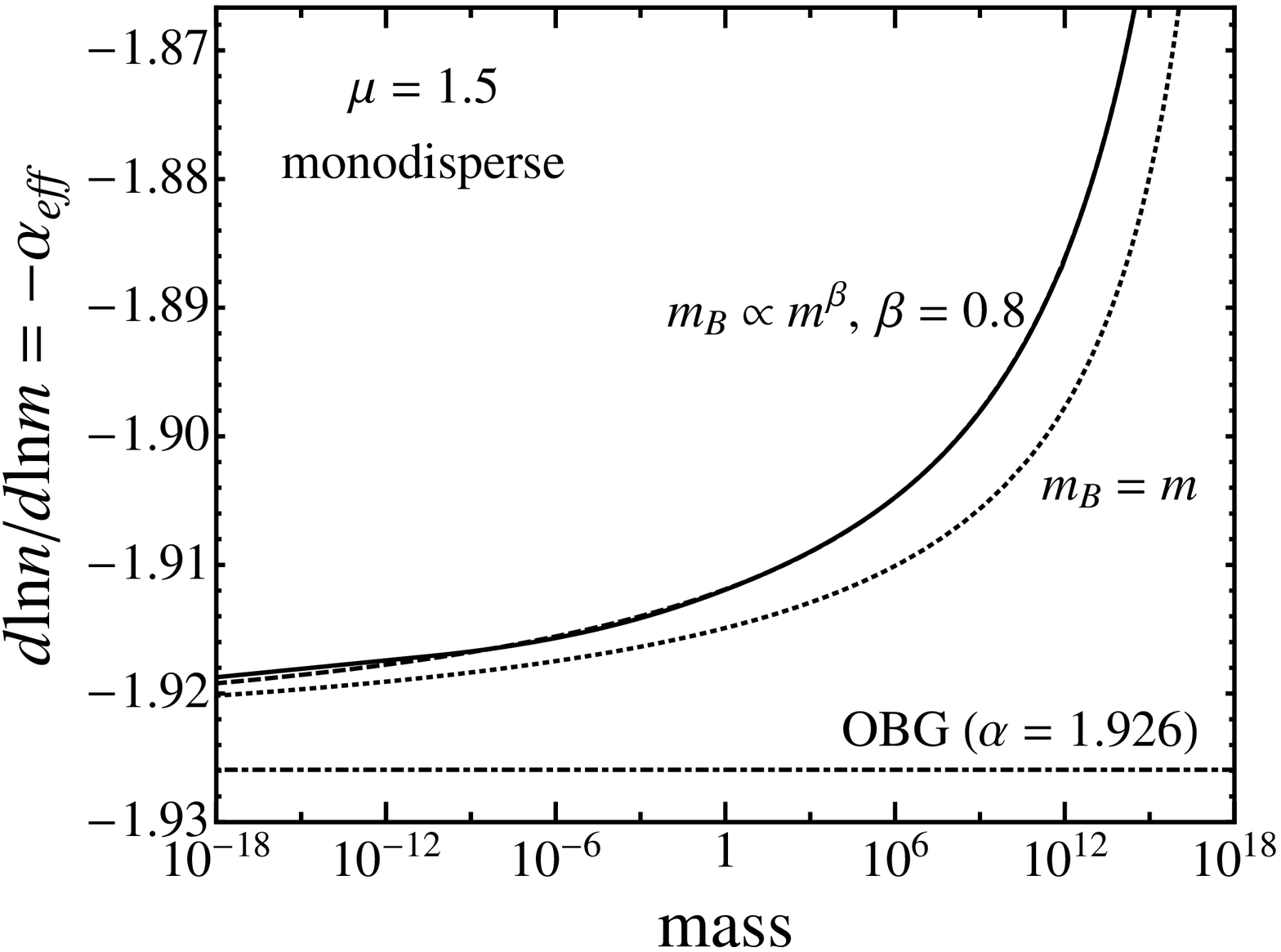}}
  \subfigure[]{\includegraphics[width=.49\textwidth]{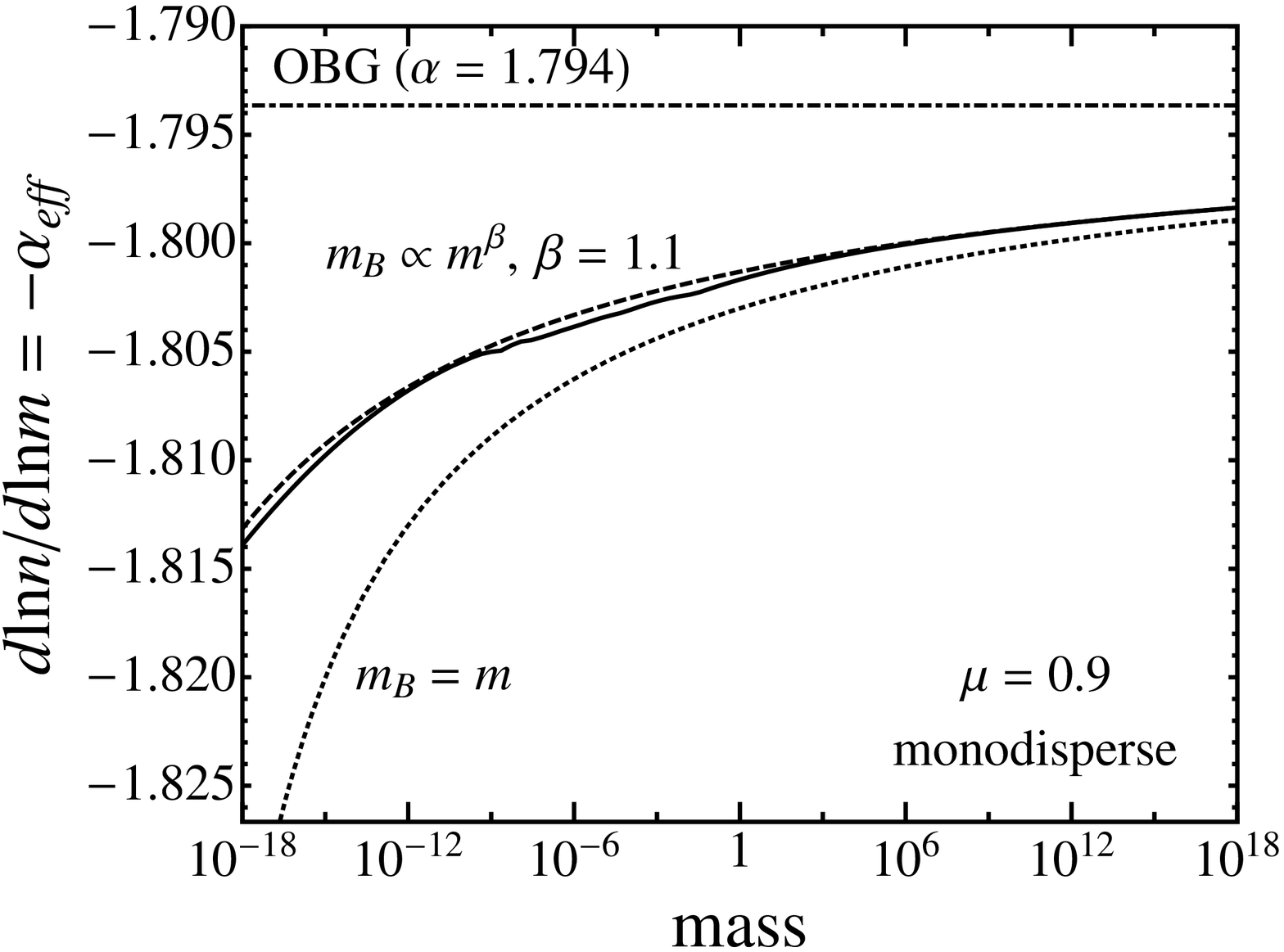}}
  \caption{Population index of the mass distribution vs. mass for the
    cases $\beta < 1, \ \mu >
1$ and $\beta > 1, \ \mu < 1$. Calculations were initialized at the analytical 
steady-state solution for each model. The dot-dashed line indicates 
the $\alpha$ given by Eq. (\ref{eq:OBG}) --- the OBG power law 
index which would be expected for the $\mu = 1$, $\beta\neq 1$ case. 
The dotted line represents the monodisperse solution with
$m_B(m) = m$. Panel (a): The analytic solution (\ref{eq:varphi_sol_3_md}) (dashed line) is
compared with the numerical result (solid line) for a monodisperse
fragmentation model with $\beta = 0.8, \ \mu = 1.5$, $m_{0,B}(m=10^{18}) =
0.1m$, and $m_B(m=10^{-18}) = 10^{-6}m$. The location of
the break is $m_{break} = 10^{4}$. Panel (b): Similar to panel (a), but
now Eq. (\ref{eq:varphi_sol_4_md}) was used to initialize the 
analytic solution for $\beta =1.1, \ \mu = 0.9$, $m_{0,B}(m=10^{-18})
    = 0.03m$, and $m_B(m=10^{18}) =10^{-4}m$. 
The location of the break is at $m_{break} = 10^{-4}$. In both panels
    (a) and (b), the numerical and analytical results are in agreement.}
 \label{mbnem_nonDohnanyi_fig}
\end{figure}

We start by looking at the case of
$\beta=1$, so we have $m_B(m)=Bm$, $B<1$. For 
a given set of parameters, we initialize $n_0(m)$ to be the analytical 
steady-state solution given by Eq. 
(\ref{eq:varphi_sol_1_norm}) if $\mu > 1$ or by Eq.
(\ref{eq:varphi_sol_2_norm}) if $\mu < 1$. 
We then check whether $n(m,t)$ evolves away 
from $n_0(m)$. If it does not, then $n_0(m)$ is the steady-state solution.

In Fig. \ref{mbnem_Dohnanyi_fig}a we compare the analytical 
formula (\ref{eq:varphi_sol_2_norm}) for $\mu=0.5$ 
with the numerical solution.
Despite small deviations between the two solutions 
(likely due to some problems with boundary conditions, see 
Appendix \ref{codeappendix}, and, possibly, weak wave 
excitation) the overall agreement between them is quite good. 
Figure \ref{mbnem_Dohnanyi_fig}b provides a comparison between our  
formula (\ref{eq:varphi_sol_1_norm}) for $\mu=1.5$ 
and the numerical solution.
In this case, the two solutions agree with each other so well
that they are hard to distinguish. 

We next look at the case of $\mu\neq 1$ and $\beta\neq 1$, so that
$m_B$ is given by Eq. (\ref{eq:mB}). Proceeding
as before, we display in Fig.
\ref{mbnem_nonDohnanyi_fig} the evolution of numerical
curves for $\alpha_{eff}$ away from analytical
solutions computed in Appendix \ref{ap:betaneq1}. 
The panels in this figure show our results for $\beta < 1, \ \mu >
1$ and $\beta > 1, \ \mu < 1$. We do not show the results for $\beta >
1, \ \mu > 1$ and $\beta < 1, \ \mu < 1$ due to numerical difficulties
with imposing boundary conditions in these two cases (Appendix
\ref{codeappendix}).
The agreement between the analytical formula 
(\ref{eq:varphi_sol_3_md}) for $\beta < 1$ and the numerical results 
displayed in Fig. \ref{mbnem_nonDohnanyi_fig}a is quite good, and
the same is true regarding the
agreement between the formula (\ref{eq:varphi_sol_4_md}) 
for $\beta > 1$ and the numerical results displayed  
in Fig. \ref{mbnem_nonDohnanyi_fig}b. 

In both Fig. \ref{mbnem_Dohnanyi_fig} and
Fig. \ref{mbnem_nonDohnanyi_fig}, we display $\alpha_{eff}$ computed for
$m_B=m$ and a monodisperse fragment size
distribution (\S\ref{sec:mbm_pow}) by a dotted line. One can see that
even though we are now using $m_B \ll m$, the solutions
are qualitatively similar to the $m_B=m$ case. Thus, one can use the
fully analytic solution
(\ref{eq:nonpl}) with $\varphi(m)$ given by Eq. (\ref{phimbm}) 
to get a qualititative picture of the non-power law behavior
regardless of the precise form of
$m_B$. We also note that the solutions for 
$\alpha_{eff}$ shown in Fig. \ref{mbnem_Dohnanyi_fig} and
Fig. \ref{mbnem_nonDohnanyi_fig} lie {\it above} the solution 
corresponding to the $m_B = m$ case. Thus, $m_B \neq m$
gives rise to a function $\varphi(m)$ which
is {\it shallower} than for the $m_B=m$ case.


\section{Discussion}
\label{discussion}



\subsection{Validity of $|d\ln \varphi/d\ln m| \ll 1$}
\label{subsect:validity}

In deriving our analytical results, we have assumed that $|d\ln
\varphi/d\ln m| \ll 1$ (Eq. \ref{eq:weak_var}). Results from the 
previous section demonstrate that the qualitative behavior of
$\varphi(m)$ is insensitive to the specific form of $m_B$. Thus we can
get a sense of when this
assumption is valid by using the form of $d\ln\varphi/d\ln
m$ for the monodisperse case with $m_B = m$: 
\ba
\frac{d\ln\varphi(m)}{d\ln m}=
-\frac{1}{2\ln(m/m_0^\star)},
\label{eq:verif}
\ea
In order to have $|d \ln \varphi/d \ln m| \ll 1$, we must have
$|\ln(m/m_0^\star)| \gg 1$, where $m_0^\star$ was defined in
\S\ref{sec:mbm_pow}. In practice, we find that even for
$|\ln(m/m_0^\star)| \sim 4$ our analytical solutions give an accurate
description of the non-power law behavior. For instance, in
Fig. \ref{powlawfig}b it is clear that for the monodisperse
fragmentation law, the exact solution for $\varphi(m)$ (Eq. (\ref{phimbm}))
works very well, even though we have $|\ln(m/m_0^\star)| = 4.6$ at $m
= 10^{-18}$.


\subsection{Comparison with existing studies}
\label{subsect:compare}

\begin{figure}[!h]
  \centering
  \includegraphics[width=.49\textwidth]{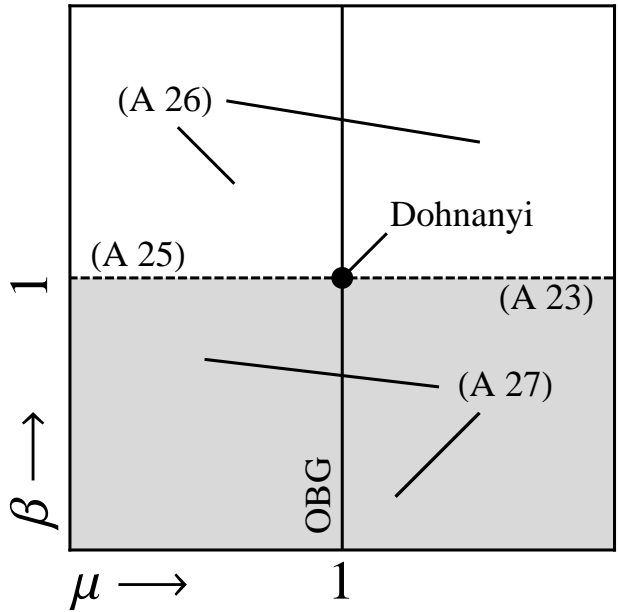}
  \caption{Parameter space plot in the $\mu-\beta$ plane. 
    The case considered by \citet{Dohnanyi,Tanaka} is at the point $\mu =
    1,  \beta=1$, and the case considered by \citet{OBG} lies on
    the line $\mu = 1$ (solid line). Our solutions cover the rest of 
    phase space and are labeled with references to corresponding 
    equations in this work. Thus, solutions (\ref{eq:varphi_sol_2_norm})
    and (\ref{eq:varphi_sol_1_norm}) 
    (monodisperse, $m_B(m) = Bm$) lie on rays $\mu < 1, \beta=1$ and 
    $\mu > 1, \beta=1$, correspondingly. 
    Solutions (\ref{eq:varphi_sol_4_md}) and 
    (\ref{eq:varphi_sol_3_md}) (monodisperse,
    $m_B(m) = Bm^\beta$) are valid in the half planes
    $\beta > 1$ (white) and $\beta < 1$ (gray) respectively. 
    For these solutions $\varphi =$const along the line $\mu = 1$ 
    in agreement with \citet{OBG}.}
 \label{parameterspace}
\end{figure}

We illustrate how our work fits into existing studies of
fragmentation cascades with a parameter space plot in $\mu-\beta$
coordinates. Figure \ref{parameterspace} shows the domains of
applicability in the $\mu - \beta$ plane for the Dohnanyi and OBG
solutions in relation to our analytic
solutions for the monodisperse case. Each of our solutions is
labeled by its corresponding formula number, and it is evident that  
our investigation covers the remainder of the $\mu-\beta$ plane.

We next discuss why previous authors have not seen non-power law
behavior. The reason is  that they have all assumed $m_{0,B}
\propto m_t$, and we have shown in \S\ref{subsect:failure} that
non-power law behavior only results 
when $m_{0,B}$ is not proportional to $m_t$. In some studies, the
assumption $m_{0,B} \propto m_t$ was explicit such as in
\citet{Dohnanyi} and \citet{OBG} who both assumed $m_0 = Cm_t$, and in
\citet{PetitFarinella}, \citet{OBG1}, and \citet{Brunini} who all
assumed\footnote{Note that $m_{0,B} = m_t/2$ does not follow
  from $m_{rem}(m_B(m_t),m_t) = m_t/2$, because the remnant does not
  belong to the distribution of ejecta (\S\ref{setup}).} $m_{0,B}
= m_t/2$. In other cases, such as
\citet{Tanaka} and
\citet{KobayashiTanaka} the scaling $m_{0,B} \propto m_t$ was implicit in
assumptions about the form of $f_{ej}$
(\S\ref{subsect:Dohnanyi},\S\ref{subsect:OBG}).

We now return to our argument from \S\ref{subsect:m0_mt} that $m_{0,B}$
should have the form (\ref{eq:m0B}). This conclusion was based on an
extension of the experimental results of \citet{Fujiwara} beyond the
strength-dominated regime. We mention that a number of authors
\citep{PetitFarinella, OBG1, Brunini} have considered a different
extension of those results, and instead of our
Eq. (\ref{eq:Fujiw}), these authors used
\ba
\frac{m_0(m_t,m_p)}{m_t} &\propto& \left(\frac{E_{coll}(m_t,m_p)/2}{Q_S(m_t)
  m_t}\right)^{-\gamma},
\label{Ereleq}
\ea
Here, $Q_S$ is the energy per unit mass required to shatter an
object, but not necessarily to disperse its fragments to
infinity \citep{OBG1}. If we assume for simplicity that $Q_D^\star
\propto Q_S$, then from
Eq. (\ref{Ereleq}) and Eq. (\ref{eq:Ecrit}), we have
$m_{0,B} \propto m_t$, for any functional form of $m_B$. The reason
it is possible to derive two different forms for $m_0(m_t,m_p)$
from the results of \citet{Fujiwara} (Eq. (\ref{eq:Fujiw}) and
Eq. (\ref{Ereleq})) is because their experiments were
performed over a small range of target masses using a constant
projectile mass. The question of how to properly
extend their results over a larger mass range can best be settled by
more experiments and simulations
\citep{Leinhardt,BenzAsphaug,BAR}, which can decisively answer how
$m_{0,B}$ varies with mass. However, we point out that unless $m_{0,B}$ is
exactly proportional to $m_t$, non-power law behavior will
result. As we show in the next section, these deviations from power
law behavior can be observationally significant
when extrapolating over many orders of magnitude in mass.


\subsection{Applications}
\label{subsect:applications}

Our results clearly demonstrate that one should be somewhat
cautious when adopting a pure power law approximation to
describe the properties of fragmentation cascades. Even though
the non-power law corrections computed in this work
scale very weakly with object mass (as the square root of
the logarithm of the mass (Eq. (\ref{phimbm}))), one has
to keep in mind that in astrophysical systems collisional
cascades span $\sim 30$ orders of magnitude in mass. Thus, even a weak
deviation from a power law can become important, such as when
inferring the disk mass (dominated by the largest bodies) from its
infrared luminosity (dominated by the smallest bodies) \citep{Wyatt}.

Just for illustration, let us consider a population of $R_{inj}=10$ 
km objects which get ground down to $R_{rm}=$1 $\mu$m size particles
by collisions. Infrared observations give us some idea of 
the mass in small particles, thus fixing the normalization
of the mass spectrum at its low-mass end, and we want to infer 
from these data the total mass in large bodies feeding 
this collisional cascade. Connecting the mass contained at the low 
and high mass ends of the spectrum by a simple power law leads
to an error caused by the neglect of the non-power law 
effects. We can estimate this error $\delta$ by using Eq. 
(\ref{phimbm}) and taking the ratio of $\varphi$ at the high and 
low mass ends. Assuming some values of $\mu$
and $C$ in Eq. (\ref{eq:m0B})  
that are ``averaged'' over the whole cascade (in practice these
parameters will
change several times between $R_{rm}$ and $R_{inj}$ because
of variations in the internal properties of objects), we have
\ba
\delta\approx \sqrt{\frac{\ln (R_{rm}/R_0^\star)}
{\ln (R_{inj}/R_0^\star)}}=\sqrt{1+\frac{\ln (R_{inj}/R_{rm})}
{\ln (R_0^\star/R_{inj})}},
\ea
where $R_0^\star$ is the radius of the object with mass $m_0^\star$
defined in \S\ref{subsect:mB=m}. Assuming for 
illustration that on ``average'' $\mu\approx 1.1$ and that at 
the high mass end the largest fragments produced in 
collisions have mass equal to $0.3$ of the 
target mass ($m_{0,B}(m_{inj})=0.3m_{inj}$) one finds
$\ln(R_0^\star/R_{inj})=\ln(m_{0,B}(m_{inj})/m_{inj})/3(\mu-1)
\approx 4$ and $\delta\approx 2-3$. Thus, in this particular 
exercise the neglect of non-power law effects
leads to an {\it underestimate} 
of the mass in large bodies by a factor of several. This also implies that
the total disk mass, which is dominated by the mass in large bodies,
is underestimated.

An underestimate of the disk mass also leads to an underestimate of
the disk lifetime, $M_{disk}/\dot{M}_{disk}$.
This occurs because if the disk is in steady-state,
then $F(m)$ is independent of $m$, which means it is possible to infer
$\dot{M}_{disk}$ from infrared observations alone, without
extrapolation to large masses \citep{Wyatt}. Supposing that we had
correctly inferred $\dot{M}_{disk}$ from observations, but had failed
to apply the non-power law correction, and hence underestimated the
disk mass, then we would also have underestimated the disk lifetime.

Based on the above discussion, breaking the assumption
$m_{0,B} \propto m_t$ affects the calculation of disk properties
from observations. Conversely, observations of disks can be used to
constrain the model parameters (i.e. $\mu$ and $C$ if $m_{0,B}$ is
given by Eq. (\ref{eq:m0B})), if e.g. the inferred disk mass is found to be
unreasonable for some parameter range. However, as mentioned in
\S\ref{subsect:compare},
direct application of our theoretical results to the observed mass 
spectrum of objects is complicated by
the multitude of additional factors playing an important role in real
astrophysical systems. Nevertheless, modern calculations
\citep{Brunini, OBG1}  of the
collisional evolution in the asteroid belt and of debris disks
\citep{Thebault, Krivov} aim for a precision of tens
of percent or less over a broad range of masses. At this level of
accuracy, the non-power law effects considered
in this work would play a significant role and should be 
taken into account.


\section{Summary}
\label{summary}


We have shown that unless $m_{0,B}(m_t) \propto m_t$, where $m_{0,B}$
is the mass of the largest fragment
produced in a collision with just enough energy to disperse half of
the target's mass, $m_t$, to infinity, a steady-state power law
solution for the
mass distribution, $n(m)$, is not possible. The non-power law behavior
is weak, however, and the 
solution for $n(m)$ becomes the product of a
power law and a much more
slowly varying function of the mass: $n(m) = m^{-\alpha}\varphi(m)$,
$|d \ln \varphi/ d \ln m| \ll 1$. This slowly varying function is
equal to a constant when $m_{0,B}(m_t) \propto m_t$, and the fact that
previous researchers \citep{KobayashiTanaka, Tanaka, Dohnanyi,
  PetitFarinella, OBG, OBG1, Williams, Brunini} have assumed just such a
dependence of $m_{0,B}(m_t)$,
explains why this kind of non-power law behavior was not observed earlier.

When $m_{0,B}(m_t)$ is not
proportional to $m_t$, $n(m)$ deviates smoothly away from a pure power
law, with the deviation only becoming significant when considering 
many orders of magnitude in mass. This is quite
different from the wavy non-power law behavior resulting
from either a lower cutoff to the mass distribution due to the ejection of
small particles by radiation pressure \citep{Thebault, CampoBagatin,
  DD}, a transition from a
collisionally-evolved to a primordial size distribution \citep{Fraser,
  PanSari, KB2004}, or a break in the power law index of the strength law
\citep{OBG, OBG1}. The non-power law behavior we describe in this work
is
significant when extrapolating over many orders of magnitude in mass,
such as when
inferring the number of large bodies in a system based on infrared
observations \citep{Wyatt}. For instance, assuming $m_{0,B}(m_t)
\propto m_t^{1.1}$, a deviation of only $10\%$ in the power law index
from the usual assumption of $m_{0,B}(m_t) \propto m_t$ results in a
factor of $\sim 2-3$ correction when inferring the number of $10$ km
bodies from observations of dust. 

We have quantified precisely the effect of the non-power law behavior
on the mass distribution by
obtaining analytical solutions for $\varphi(m)$ in the case
of a power law fragment mass distribution with $m_B(m_t) = m_t$, and
a monodisperse fragment mass distribution (all fragments the same
size) with $m_B(m_t) = Bm_t^\beta$.
We have also provided a general framework for solving the $m_B(m_t) =
Bm_t^\beta$ case
with an arbitrary fragment mass distribution. In all cases considered,
our analytical solutions were confirmed numerically, and we noticed
that the simple expression (\ref{phimbm}) captures the essence of the
non-power law behavior for a wide range of parameters in our model. 

In the course of our investigation, we have also found an entirely
different type of non-power law behavior. Namely, we have discovered
that fragmentation cascades can support wavy, steady-state solutions, 
even when there is
no upper or lower mass cutoff, and the strength law is given by a
pure power law. Our results were derived for the monodisperse
case, but such waves may also be able to persist for
more realistic fragment mass distributions. In astrophysical systems,
these kinds of waves could be triggered in stochastic collisions between large
planetesimals that generate enough collisional debris to significantly
alter $n(m)$.

We are grateful to Jeremy Goodman for useful discussions. 
The financial support for this work is provided by the Sloan 
Foundation and NASA via grant NNX08AH87G.

\appendix

\section{Convergence of the Mass Flux Integral}
\label{subsect:converge}

We discuss here the convergence of the mass flux integrals in
\S\ref{subsect:Dohnanyi},\ref{subsect:OBG}. If the correct value of
$\alpha$ is substituted into
Eq. (\ref{eq:flux_doh}) or Eq. (\ref{eq:flux_obg}), then the mass flux
takes the form
\ba
F(m)\propto \int\limits_0^1 dx x^{-1} \int\limits_0^\infty
dy y^{-\alpha} f(x,y).
\label{eq:flux_powlaw}
\ea
It is helpful to discuss under what circumstances the integrals
in this expression converge. 

The integral over $y$ converges at its upper limit if
$\alpha > 1$, because $f(x,y) < 1$. There is actually
already a more stringent condition of $\alpha > 5/3$, which comes from
the requirement that the
cross-section be dominated by the smallest particles
(\S\ref{simplifications}), so the requirement $\alpha > 1$ is
automatically fulfilled.

We now consider convergence of the integral over $y$ at its lower
limit. Generally speaking, the quantity
$f(x,y)y^{-\alpha}$ drops off for $y \ll 1$, the threshold for
catastrophic breaking, which leads to convergence. In
fact, in a model without erosion, $f(x,y) = 0$ for $y < 1$, so there
is a sharp cutoff at $y=1$. In reality, erosion will smooth this
cutoff, but as long as $f(x,y)$ falls off faster than $y^{\alpha-1}$
for $y \ll 1$, then the integral will converge at its lower limit. We
point out that if the integral over $y$ converges, then it follows
from the above arguments that
$f(x,y)y^{-\alpha}$ is peaked at $y_{peak}=k$. In the absence of erosion, it
is clear that $k \sim 1$, but when erosion is considered, we
typically have $k \ll 1$ \citep{KobayashiTanaka}. 

We next consider the convergence of the integral over $x$, and we
find it helpful to define 
\ba
f(x) \equiv \int_0^\infty dy y^{-\alpha} f(x,y)
\label{eq:fx}
\ea
As long as $f(x)$ is bounded on $x \in [0,1]$, then the integral
over $x$ in Eq. (\ref{eq:flux_powlaw}) converges at the upper
limit of integration. At the lower limit of integration, the function
$x^{-1}$ diverges, but only logarithmically, so if $f(x) \rightarrow
0$ for $x \ll 1$, then we would typically expect convergence at
the lower limit. From Eq. (\ref{eq:gdef}) and Eq. (\ref{eq:x}), we have
$f(x,y) \rightarrow 0$ for $x \ll 1$, so we do indeed expect
$f(x) \rightarrow 0$ in the same limit.


\section{Details of the calculation of $\varphi(m)$.}  
\label{app:varphi}


We start by developing a general method for calculating 
$\varphi(m)$ for different forms of $m_B(m)$ from the 
functional Eq. (\ref{eq:Theta}) with known $\Theta(m)$.

\subsection{Case $m_B(m) = Bm$, $\beta=1$}

Here, we will first assume following Dohnanyi (1969) and  Tanaka \etal 
(1996) that $m_B(m)=Bm$. Then, we need to solve the functional equation
\ba
\varphi(m)\varphi\left(Bm\right)=\Theta(m).
\ea
Taking the logarithm of both sides of 
this equation we get
\ba
\ln\varphi(m)+\ln\varphi\left(Bm\right)=\ln\Theta(m),
\label{eq:logarithmized_1}
\ea
and upon introducing the new independent variable 
$v\equiv\ln m$, the constant $b\equiv\ln B$, and the new function 
$\varphi_1(s)\equiv \ln\varphi\left(e^s\right)$ one gets
\ba
\varphi_1(v)+\varphi_1(v+b)=R_1(v),~~~R_1(v)\equiv 
\ln\Theta(e^v).
\label{eq:varphi_1}
\ea
For some applications, it is more appropriate to study an equivalent 
equation
\ba
\varphi_1(u-b)+\varphi_1(u)=R_1(u-b).
\label{eq:varphi_2}
\ea

One can check by direct substitution that the formal solution of 
Eq. (\ref{eq:varphi_1}) up to an additive constant
$\varphi_1(-\infty)$ is given by
\ba
\varphi_1(v) + \varphi_1(-\infty) = \sum\limits_{k=0}^\infty (-1)^k R_2(v+kb),
\label{eq:varphi_1_sol}
\ea
where we have used the fact that $b < 0$, since $B < 1$. It follows
then, that $\varphi$ is given up to an overall 
normalization constant $\varphi(0)$ as
\ba
\frac{\varphi(m)}{\varphi(0)}=\exp\left[\sum\limits_{k=0}^\infty (-1)^{k} 
\ln\Theta(e^{\ln m + kb})\right]=\prod\limits_{k=0}^\infty
\frac{\Theta\left(me^{2kb}\right)}{\Theta\left(me^{(2k+1)b}\right)}.
\label{eq:varphi_sol_1}
\ea
Analogously, the solution of Eq. (\ref{eq:varphi_2}) up to an
additive constant and the corresponding solution for $\varphi$ up to a
normalization are given by
\ba
&& \varphi_1(v)+ \varphi_1(\infty)=\sum\limits_{k=0}^\infty (-1)^k 
R_1(v-(k+1)b),
\label{eq:varphi_2_sol}\\
&& \frac{\varphi(m)}{\varphi(\infty)}=\prod\limits_{k=0}^\infty
\frac{\Theta\left(me^{-(2k+1)b}\right)}{\Theta\left(me^{-(2k+2)b}
\right)}.
\label{eq:varphi_sol_2}
\ea
In \S \ref{ap:betaeq1} we provide an example of how to
discriminate between using Eq. (\ref{eq:varphi_sol_1}) versus
Eq. (\ref{eq:varphi_sol_2}) to calculate $\varphi$. 


\subsection{Case $m_B(m) = Bm^{\beta}, \ \beta \ne 1$}

Whenever $\beta \neq 1$ we need to solve the functional Eq.
$\varphi(m)\varphi\left(Bm^\beta\right)=\Theta(m)$, 
which is easily converted to 
\ba
\ln\varphi(m)+\ln\varphi\left(Bm^\beta\right)=\ln\Theta(m).
\label{eq:logarithmized_2}
\ea
Invoking the mass scale $m_B^\star\equiv B^{1/(1-\beta)}$ at which
breaking stops (i.e. $m_B(m_B^\star) = m_B^\star$),
and defining the new independent variable 
$u\equiv \ln|\ln(m/m_B^\star)|$, $-\infty<u<\infty$ 
and the new function $\varphi_2(s)\equiv
\ln\varphi\left(m_B^\star e^{e^s}\right)$ one obtains
\ba
\varphi_2(u)+\varphi_2(u+a)=R_2(u),~~~R_2(u)\equiv \ln
\Theta\left(m_B^\star e^{e^u}\right),~~~a\equiv \ln\beta.
\label{eq:varphi_3}
\ea
This equation can also be converted to an equivalent form useful
in some applications:
\ba
\varphi_2(u-a)+\varphi_2(u)=R_2(u-a).
\label{eq:varphi_4}
\ea

Analogous to the previous case, we can write down formal solutions of 
Eq. (\ref{eq:varphi_3}) and Eq. (\ref{eq:varphi_4}) up to
additive constants, and the corresponding solutions for $\varphi$ up to
normalization constants. For $\beta > 1$, we have $a > 0$, which yields
for Eq. (\ref{eq:varphi_3}) 
\ba
&& \varphi_2(u) + \varphi_2(\infty)=\sum\limits_{k=0}^\infty (-1)^k R_2(u+ka),
\label{eq:gen_sol}\\
&& \frac{\varphi(m)}{\varphi(0)}=\prod\limits_{k=0}^\infty
\frac{\Theta\left(m_B^\star\exp\left(e^{2ka}\ln\frac{m}{m_B^\star}\right)\right)}
{\Theta\left(m_B^\star\exp\left(e^{(2k+1)a}\ln\frac{m}{m_B^\star}\right)\right)},
\label{eq:varphi_sol_3}
\ea
and for Eq. (\ref{eq:varphi_4}) 
\ba
&& \varphi_2(u)+\varphi_2(-\infty)=\sum\limits_{k=0}^\infty (-1)^k R_2(u-(k+1)a),
\label{eq:varphi_4_sol}\\
&& \frac{\varphi(m)}{\varphi(m_B^\star)}=\prod\limits_{k=0}^\infty
\frac{\Theta\left(m_B^\star\exp\left(e^{-(2k+1)a}\ln\frac{m}{m_B^\star}\right)\right)}
{\Theta\left(m_B^\star\exp\left(e^{(-2k+2)a}\ln\frac{m}{m_B^\star}\right)\right)}.
\label{eq:varphi_sol_4}
\ea

In a similar fashion, we can obtain the solutions when $\beta < 1$,
in which case $a < 0$. We have for Eq. (\ref{eq:varphi_3})
\ba
&& \varphi_2(u) + \varphi_2(-\infty)=\sum\limits_{k=0}^\infty (-1)^k R_2(u+ka),
\label{eq:gen_sol_a}\\
&& \frac{\varphi(m)}{\varphi(m_B^\star)}=\prod\limits_{k=0}^\infty
\frac{\Theta\left(m_B^\star\exp\left(e^{2ka}\ln\frac{m}{m_B^\star}\right)\right)}
{\Theta\left(m_B^\star\exp\left(e^{(2k+1)a}\ln\frac{m}{m_B^\star}\right)\right)},
\label{eq:varphi_sol_3_a}
\ea
and for Eq. (\ref{eq:varphi_4})
\ba
&& \varphi_2(u)+\varphi_2(\infty)=\sum\limits_{k=0}^\infty (-1)^k R_2(u-(k+1)a),
\label{eq:varphi_4_sol_a}\\
&& \frac{\varphi(m)}{\varphi(\infty)}=\prod\limits_{k=0}^\infty
\frac{\Theta\left(m_B^\star\exp\left(e^{-(2k+1)a}\ln\frac{m}{m_B^\star}\right)\right)}
{\Theta\left(m_B^\star\exp\left(e^{(-2k+2)a}\ln\frac{m}{m_B^\star}\right)\right)}.
\label{eq:varphi_sol_4_a}
\ea

This completes the description of the general mathematical 
formalism needed for finding $\varphi(m)$ given $\Theta(m)$ 
and given $m_B(m)$ in power law form. We now obtain explicit 
expressions for $\varphi(m)$ for a monodisperse fragment 
mass distribution.


\subsection{Application to the monodisperse case}  
\label{app:monodisp}


\subsubsection{Case $m_B(m) = Bm^\beta$, $\beta=1$}
\label{ap:betaeq1}
In the monodisperse case, we can take
\ba
\Theta(m)=\frac{1}{\ln(m/m_{0,B}(m))} = \frac{1}{(1-\mu)\ln m-\ln C},
\label{eq:theta}
\ea
where we will again assume $m_{0,B}(m) = Cm^\mu$. Considering the case
$\beta = 1$, the
two relevant equations for obtaining $\varphi(m)$ are
Eq. (\ref{eq:varphi_sol_1}) and
Eq. (\ref{eq:varphi_sol_2}). The $\varphi(0)$ term in the denominator on
the left hand side of Eq. (\ref{eq:varphi_sol_1})
means that this equation is only applicable when $\mu > 1$, since for $\mu <
1$ there is a value $m_0^\star$ below which $m_{0,B}(m) > m$, and the
solution is unphysical below this
point. Similarly, the $\varphi(\infty)$ term in the denominator on
the left hand side of Eq. (\ref{eq:varphi_sol_2}) means that the
solution only works for $\mu < 1$, because for
$\mu > 1$, $m_{0,B}(m) > m$ above $m_0^\star$, and the solution is
again unphysical. 

Treating first the $\mu > 1$ case, if we substitute Eq.
(\ref{eq:theta}) into Eq. (\ref{eq:varphi_sol_1}), we have 
\begin{equation}
\label{eq:varphi_sol_1b}
\frac{\varphi(m)}{\varphi(0)} = \prod_{k=0}^\infty
\frac{\ln\left(\frac{m}{m_{0,B}(m)}\right)+(2k+1)(1-\mu)b}
{\ln\left(\frac{m}{m_{0,B}(m)}\right)+2k(1-\mu)b}.
\end{equation}
Unfortunately, this expression does not converge, implying that
$\varphi(0)  = 0$. To understand this behavior, we
can refer back to the analytic solution for $\varphi$ in the monodisperse
case with $m_B(m) = m$ (Eq. (\ref{phimbm})). There, we had $\varphi(m)
= 1/\sqrt{\ln(m/m_{0,B}(m))}$. For $\mu > 1$, this expression does indeed
yield $\varphi(0) = 0$. 

Since $\varphi(m)$ is only defined up to a normalization, we can remedy 
the situation by working with the 
{\it ratio} of $\varphi(m)$ at two points rather than with
$\varphi(m)$ itself, which gives a convergent expression. 
If we define $a_1
\equiv \ln (m_1/m_{0,B}(m_1))$, $a_2 \equiv \ln (m_2/m_{0,B}(m_2))$, and $c
\equiv (1-\mu)b$, then it follows from expression
(\ref{eq:varphi_sol_1b}) that 
\begin{equation}
\frac{\varphi(m_1)}{\varphi(m_2)} = \prod_{k=0}^\infty \frac{(a_1 +
  (2k+1)b)(a_2 + 2kb)}{(a_1 + 2kb)(a_2 + (2k+1)b)},
\end{equation}
which simplifies to
\begin{equation}
\label{eq:varphi_sol_1_norm}
\frac{\varphi(m_1)}{\varphi(m_2)} = \frac{a_2 \Gamma
  \left(1+\frac{a_1}{2c}\right) \Gamma \left(\frac{1}{2} +
  \frac{a_2}{2c}\right)}{a_1 \Gamma \left(1 + \frac{a_2}{2c} \right)
  \Gamma\left(\frac{1}{2} + \frac{a_1}{2c}\right)}. 
\end{equation}
The correctness of this solution can be verified
directly by substituting $\varphi$ back into Eq. 
(\ref{eq:Theta}), and using
(\ref{eq:theta}) for $\Theta(m)$.

Treating next the $\mu < 1$ case, if we substitute Eq.
(\ref{eq:theta}) into Eq. (\ref{eq:varphi_sol_2}), we have
\ba
\label{eq:varphi_sol_2_md}
\frac{\varphi(m)}{\varphi(\infty)}=\prod_{k=0}^\infty
\frac{\ln\left(\frac{m}{m_{0,B}(m)}\right)-(2k+2)(1-\mu)b}
{\ln\left(\frac{m}{m_{0,B}(m)}\right)-(2k+1)(1-\mu)b}.
\ea
Again, this equation does not converge, but the ratio of $\varphi$ at
two points does, and using the same definitions for $a_1$, $a_2$, and
$c$ as before, we have
\begin{equation}
\label{eq:varphi_sol_2_norm}
\frac{\varphi(m_1)}{\varphi(m_2)} = \frac{\Gamma
  \left(1-\frac{a_2}{2c}\right) \Gamma \left(\frac{1}{2} -
  \frac{a_1}{2c}\right)}{\Gamma \left(1 - \frac{a_1}{2c} \right)
  \Gamma\left(\frac{1}{2} - \frac{a_2}{2c}\right)}.
\end{equation}


\subsubsection{Case $m_B(m) = Bm^{\beta}, \ \beta \ne 1$}
\label{ap:betaneq1}

We now treat the case when the mass of the smallest projectile that
can catastrophically shatter a target is not proportional to the mass
of the target itself, so that $\beta \neq 1$. 

We first consider the case $\beta > 1$, so we have the choice of using
Eq. (\ref{eq:varphi_sol_3}) or Eq. (\ref{eq:varphi_sol_4}),
and we
limit ourselves to the situation when $m_{0,B}(m_B^\star) < m_B^\star$. We
expect such a situation to be realistic, since the
maximum fragments created by the disruption of bodies with mass
close to $m_B^\star$, should still be smaller than $m_B^\star$. 
Without this assumption, $\varphi(m_B^\star)$
would be unphysical, since we would have $m_{0,B}(m_B^\star) > m_B^\star$,
which is impossible in a fragmentation cascade. Given this
assumption, we can use Eq. (\ref{eq:varphi_sol_4}) to solve for
$\varphi$ up to a normalization.

Substituting
Eq. (\ref{eq:theta}) into Eq. (\ref{eq:varphi_sol_4}), we have
\ba
\frac{\varphi(m)}{\varphi(m_B^\star)}=\prod\limits_{k=0}^\infty\frac{
\ln\left(\frac{m}{Dm_{0,B}(m)}\right)\beta^{-(2k+2)}+
\ln D}
{\ln\left(\frac{m}{Dm_{0,B}(m)}\right)\beta^{-(2k+1)}+
\ln D},
\label{eq:varphi_sol_4_md}
\ea
where we have defined the constant $D \equiv
C^{-1}B^{(1-\mu)/(1-\beta)} = (m_B^\star/m_0^\star)^{1-\mu}$. In this
case, the product does converge since $\varphi(m_B^\star)$ is nonzero (it
is also finite), and we do not have to take the ratio of
two points as we did for the $\beta = 0$ case in \S \ref{ap:betaeq1}.

Next, we consider the case $\beta < 1$, and now we have the choice of
using Eq. (\ref{eq:varphi_sol_3_a}) or Eq.
(\ref{eq:varphi_sol_4_a}). We again make the assumption that $m_{0,B}(m_B^\star)
< m_B^\star$, in which case we can use Eq. (\ref{eq:varphi_sol_3_a}) to obtain 
\ba
\frac{\varphi(m)}{\varphi(m_B^\star)}=\prod\limits_{k=0}^\infty\frac{
\ln\left(\frac{m}{D m_{0,B}(m)}\right)\beta^{2k+1}+
\ln D}
{\ln\left(\frac{m}{Dm_{0,B}(m)}\right)\beta^{2k}+
\ln D}
\label{eq:varphi_sol_3_md}.
\ea
Again, there are no problems with convergence. Note, that if 
we set $\mu=1$ then $m/m_{0,B}(m)=$const and Eq.
(\ref{eq:varphi_sol_4_md}) and Eq. (\ref{eq:varphi_sol_3_md}) 
yield $\varphi(m)=$const in agreement with \citet{OBG}.


\section{Fragmentation Code}
\label{codeappendix}

We describe here the numerical algorithm we use to study
fragmentation. We differ from the main text here in that our 
algorithm evolves the differential number density
of particles $n(r)$ per unit radius, rather than particle
mass, but it is straightforward to convert between $n(m)$ and $n(r)$. 

The evolution equation for $n(r)$ can be written as the sum of a
source term (denoted by a plus sign) and a sink term (denoted by a
minus sign):
\begin{equation}
\frac{\partial n}{\partial t}(r,t) = \frac{\partial n_-}{\partial
  t}(r,t) + \frac{\partial n_+}{\partial t}(r,t). 
\end{equation}
The sink term is simply given by the number of catastrophic collisions
that particles with radius $r$ are undergoing per unit time. Dropping
the dependence on time for brevity (everything is evaluated at time
$t$) and ignoring erosion, we can write
\begin{equation}
\label{sink}
\frac{\partial n_-}{\partial t}(r) = - \pi v n(r) \int_{r_B(r)}^{\tilde 
r_B(r)}
dr' n(r') (r+r')^2,
\end{equation}
where $r_B(r)$ is the minimum particle size that can fragment a
particle of radius $r$, $\tilde r_B(r)$ is defined analogously to  
$\tilde m_B(m)$, and we have assumed that the particle cross-section is
the geometric cross-section, and that the impact velocity is a constant
\S\ref{simplifications}.

For the source term, we will first state the equation and then analyze it
piece by piece:
\begin{equation}
\label{source}
\frac{\partial n_+(r)}{\partial t} = \pi v \int_r^\infty d r_1 n(r_1)
\int_{r_B(r_1)}^{\tilde r_B(r_1)} dr_2 n(r_2) h(r, r_1, r_2) (r_1+r_2)^2 .
\end{equation}
Here, we have used $r_1$ to denote a target and $r_2$ a
projectile with the possibility that $r_2 > r_1$. Now,
$\pi v n(r_1)n(r_2)(r_1+r_2)^2dr_2dr_1$ gives the number of collisions (per
unit volume per unit time)
between targets in the range  $r_1$ to $r_1 + dr_1$ and projectiles in
the range
$r_2$ to $r_2 + dr_2$. We define $h(r, r_1, r_2)dr$ to be the number of
fragments in the size range $r$ to $r + dr$ from destruction of the target
  particle only. Then $h(r, r_1, r_2)
n(r_1)n(r_2){\cal R}(r_1,r_2)dr_2dr_1 dr$ gives the particle flux into
the range $r$ to $r+dr$ from targets in the range
$r_1$ to $r_1 + dr_1$ that have been shattered by projectiles in the
range $r_2$ to $r_2 + dr_2$. Then, we
simply integrate this expression over
all values of $r_1 > r$ and all values of $r_2$, which yield a
catastrophic fragmentation event (i.e. an event in which both the
target and the projectile are destroyed).

A useful simplification of Eq. (\ref{source}) can be obtained if we
assume that the distribution of fragments is independent of the
projectile size so that $h(r, r_1, r_2) \rightarrow h(r, r_1)$. In
this case, which is adopted for the numerical calculations in this
work (\S\ref{numerical}), equation (\ref{source}) becomes
\begin{equation}
\frac{\partial n_+}{\partial t}(r) = \pi v \int_r^\infty dr_1 
h(r, r_1) n(r_1)
\int_{r_B(r_1)}^{\tilde r_B(r_1)} dr_2 n(r_2) (r_1+r_2)^2,
\end{equation}
and comparing with Eq. \ref{sink} we see that we can write 
\begin{equation}
\label{sourcespecial}
\frac{\partial n_+}{\partial t}(r) = - \int_r^\infty dr' h(r,r')
\frac{\partial n_-}{\partial t}(r').
\end{equation}
This is a useful form of the source equation when it comes to
computations, because it reduces a double integral to a single
integral, once the sink term has been calculated.

Given a starting distribution for $n(r)$ and a form for $h(r,r')$
(e.g. monodisperse, power law, etc.), it is possible to evolve the
distribution forward in time using discretized versions of Eq.
(\ref{sink}) and Eq. (\ref{sourcespecial}) in $\log r$ space. 
The integrals in these equations are performed using standard numerical
integration techniques, such as the trapezoid or Simpson's
rule. This results in an efficient order $O(N)$ method, where $N$ is the number
of radius bins which are equally spaced in $\log r$. 
The scaling of the method is important for us,
since to resolve the non-power law behavior we use up to $\sim 3000$ bins 
per decade in $r$, yielding a total of $\sim 10^5$ bins.

In order to advance the distribution in time, a timestep must be
specified.
A good criterion is to set it to a fixed fraction of the shortest collision
time in the simulation. This ensures that the particle size
distribution can never become negative. Another problem is dealing
with the boundaries of the simulation. The upper boundary is generally
not problematic since the collision time there is long compared to the
rest of the simulation, and the maximum particle size can be chosen to
be as large as necessary for there to be negligible variation in the
particle mass distribution at the high mass end. The lower boundary,
however, can be a problem if $r_B(r) \ll r$ there. In this case,
extrapolation of the particle mass distribution to lower masses is
necessary, and care must be taken to ensure that the simulation is
stable.  

\end{document}